\documentclass[traditabstract]{aa} 
\usepackage{graphicx}
\usepackage{txfonts}

\newcommand{\myrule}{\rule[-0.1cm]{0.cm}{0.7cm}} 
\newcommand\msun{M_{\odot}}
\newcommand\rsun{R_{\odot}}
\newcommand\mjup{M_\mathrm{Jup}}

\newcommand\chaha{Cha\,H$\alpha$\,}

\begin{document}
   \title{Binary frequency of very young brown dwarfs \\ at 
separations smaller than 3\,AU\thanks{Based on observations collected at the 
       European Southern Observatory, Chile 
       in program 75.C-0851(C),   
       76.C-0847(A),              
       77.C-0831(A+D).            
                    }}

   \author{V. Joergens
          \inst{1}
          }

   \institute{Max-Planck Institut f\"ur Astronomie, 
     K\"onigstuhl~17, D-69117 Heidelberg, Germany\\
     \email{viki@mpia.de}
             }

   \date{Received June 18, 2008; accepted}

{} 
 
  \abstract
   {Searches for companions of brown dwarfs 
     by direct imaging mainly probe orbital separations 
     greater than 3--10\,AU. On the other hand, 
     previous radial velocity surveys of brown dwarfs are mainly sensitive to 
     separations smaller than 0.6\,AU.
     It has been speculated that the peak of the separation distribution of brown dwarf
     binaries lies right in the unprobed range.
     This work extends high-precision radial velocity surveys of
     brown dwarfs for the first time out to 3\,AU.
Based on more than six years UVES/VLT spectroscopy 
     the binary frequency of brown dwarfs and (very) low-mass stars
     (M4.25-M8) in Chamaeleon\,I was determined:
     18$^{+20}_{-12}$\,\% for the whole sample 
     and 10$^{+18}_{-8}$\,\% for the subsample of ten brown dwarfs and very low-mass stars 
     (M$\lesssim 0.1\,\msun$).
     Two spectroscopic binaries were confirmed,
     the brown dwarf candidate \chaha8 
     (previously discovered by Joergens \& M\"uller)
     and the low-mass star CHXR\,74. 
     Since their orbital separations appear to be 1\,AU or greater,
     the binary frequency at $<$1\,AU might be less than 10\%.
Now for the first time companion searches of (young) brown dwarfs cover the 
     whole orbital separation range, and the following observational constraints for 
     models of brown dwarf formation can be derived:
     (i) the frequency of brown dwarf and very low-mass stellar binaries at $<$3\,AU 
     does not significantly exceed that at $>$3\,AU;
     i.e. direct imaging surveys do not miss a significant fraction of brown dwarf binaries;
     (ii) the overall binary frequency of brown dwarfs and very low-mass
     stars is 10-30\,\%;
     (iii) the decline in the separation distribution of brown dwarfs 
     towards smaller separations seems to occur between 1 and 3\,AU;
     (iv) the observed continuous decrease in the binary frequency from the stellar 
     to the substellar regime is confirmed at $<$3\,AU 
     providing further evidence of
     a continuous formation mechanism from 
     low-mass stars to brown dwarfs.
   }


\titlerunning{Binary frequency of very young brown dwarfs at 
separations $<$3\,AU}
\authorrunning{Joergens}

   \maketitle
%

\section{Introduction}

Searching for companions to brown dwarfs (BDs)
is essential for understanding BD formation, for which
no widely accepted model exits 
(e.g. recent review by Luhman et al. 2007).
A priori, one might expect that BDs form in
the same manner as stars by the direct collapse and fragmentation of molecular clouds, 
only on a much smaller scale.
Standard cloud fragmentation seems to have difficulty
in making BDs (Boss 2001; Bate, Bonnell \& Bromm  2003).
One possible explanation is that the simulations lack an important
piece of physics; e.g., supersonic magneto-turbulence can create local high densities
and facilitate the formation of BDs (Padoan \& Nordlund 2004).
Alternatively, 
BDs are formed when cloud fragmentation is modified by an additional process that 
prematurely halts accretion during the protostellar stage, 
such as dynamical ejection out of the dense gaseous environment
(e.g. Reipurth \& Clarke 2001; Boss 2001; Bate, Bonnell \& Bromm 2002; 
Sterzik \& Durisen 2003; Umbreit et al. 2005)
or photoevaporation by ionizing radiation from nearby massive stars 
(Kroupa \& Bouvier 2003; Whitworth \& Zinnecker 2004).
A fourth, recently studied scenario foresees BD formation through instabilities 
in the outer ($\gtrsim$100\,AU) 
disk around a star and release into the field by e.g. secular perturbation
(Goodwin \& Whitworth 2007; Stamatellos et al. 2007).

The frequency and properties of BDs in multiple systems are fundamental parameters
in these models, e.g. embryo-ejection scenarios for BD formation
predict only a few BD binaries in preferentially close orbits, while 
isolated fragmentation scenarios are expected to generate BD binaries with similar
(scaled-down) properties, as stellar binaries have.
However, the binary properties of BDs are poorly constrained for close separations.
Most of the current surveys for companions
to BDs and very low-mass stars (VLMS, $M \leq 0.1\,\msun$)
are done by direct (adaptive optics or HST) imaging 
(e.g. Reid et al. 2001; Bouy et al. 2003; Burgasser et al. 2003; Gizis et al. 2003;
Close et al. 2003; see also Burgasser et al. 2007)
and are not sensitive to either close separations
($a\lesssim2$--3\,AU and $a\lesssim10$--15\,AU for the field and clusters, 
respectively) nor to low-mass ratio systems (q$\equiv$M$_2$/M$_1\lesssim$0.6).
These surveys find a lower frequency (10-30\%) of 
BD/VLM binaries compared to stars, 
a separation distribution with a peak around 3--10\,AU,
and a preference for equal-mass components
with 77\% having a mass ratio q$\geq$0.8.
The observed peak of the separation distribution is close to the incompleteness limit;
thus, we
do not know if there is a significant number of still undetected, very close ($<$3\,AU)
BD/VLM binaries.
It has been suggested that BD/VLM binaries with a separation $<$3\,AU
are as frequent or even more frequent than those at $>$3\,AU 
(Pinfield et al 2003; Maxted \& Jeffries 2005; Burgasser et al. 2007).
Hence,
the peak of the BD/VLM binary separation distribution could lie below 3\,AU.
Furthermore, while it has been argued that the higher frequency of systems with mass ratios close
to unity appears to be a real feature of very low-mass binaries, 
it would be, nevertheless, desirable to actually probe
mass ratios lower than 0.6.
Given these limits to the current observational data, 
the question arises as to whether our current picture of BD/VLM binaries is complete,
or, whether we miss a significant fraction of very close and/or
low-mass ratio systems.

Spectroscopic monitoring of radial velocity (RV) variations provides a means to detect 
very close binaries. 
Given sufficient RV precision, the detection of
low-mass ratio systems is feasible down to planetary masses (cf. Joergens 2006a;
Joergens \& M\"uller 2007).
While several spectroscopic surveys for companions of 
BD/VLMS in young clusters
(Joergens \& Guenther 2001; 
Kenyon et al. 2005; Joergens 2006a; Kurosawa, Harries \& Littlefair 2006; 
Prato 2007a; Maxted et al. 2008)
and in the field
(Reid et al. 2002; Guenther \& Wuchterl 2003; Basri \& Reiners 2006)
have been started in recent years, data sampling is sparse in most cases and 
the number of confirmed close companions to BD/VLMS is still small.
To date, there are four spectroscopic BD/VLM binaries known for which an
orbital solution has been determined: \emph{PPl\,15} (Basri \& Martin 1999), 
\emph{GJ\,569Bab} (Zapatero Osorio et al 2004; Simon, Bender \& Prato 2006),
\emph{2MASS~J05352184-0546085} (Stassun, Mathieu \& Valenti 2006), and 
\emph{\chaha8} (Joergens \& M\"uller 2007).
One of them, \chaha8, was discovered (Joergens \& M\"uller 2007) within an RV survey for 
(planetary and BD) 
companions of very young BD/VLMS in the Chamaeleon\,I star-forming region, 
which is being carried out 
with UVES at the VLT (Joergens \& Guenther 2001; Joergens 2006a).
\chaha8 is presumably the lowest mass RV companion detected so far in a close orbit around a BD/VLMS
and only the second very young BD/VLM spectroscopic binary.
That survey in Cha\,I
provided the first determination of the binary frequency of very young BD/VLMS
at close separations and the indication that the fraction of 
undetected BD binaries at close separations must be small.
These results were based on probing orbital separations below 0.3\,AU.
The present paper reports follow-up spectroscopic observations and RV measurements
of the targets of that survey. This significantly enhances the
separation range probed. 
For the first time the binary frequency of (very young) BD/VLMS at separations of 3\,AU
and smaller is determined.

The paper is organized as follows. Section\,\ref{sect:rvs} describes
the spectroscopic observations and RV measurements,
in Sect.\,\ref{sect:RVvari} the identification of RV variable objects
is presented, in Sect.\,\ref{sect:binfrac}
the observed binary fraction is derived and the probed separation range
assessed, Sect.\,\ref{sect:chxr74} gives details for
the spectroscopic system CHXR\,74, and finally, 
Sect.\,\ref{sect:disc} and Sect.\,\ref{sect:concl}
provide a discussion and conclusions.


\section{Observations and radial velocities}
\label{sect:rvs}

Spectroscopic follow-up observations of BDs and 
(very) low-mass stars (VLMSs) in Cha\,I
were carried out between October 2005 and January 2006 with
the Uv-Visual Echelle Spectrograph (UVES) attached to the VLT 8.2\,m KUEYEN telescope
at a spectral resolution $\lambda$/$\Delta \lambda$ of 40\,000 in the 
red optical wavelength regime.
In program 76.C-0847(A), UVES spectra were taken of all objects
that were previously observed by Joergens (2006a)
and that still lacked re-observation after more than one year
(\chaha1, \chaha2, \chaha3, \chaha5,
\chaha6, \chaha12\footnote{Simbad names: [NC98]~Cha~HA~1, [NC98]~Cha~HA~2, etc.}, 
and B\,34\footnote{Simbad name: CHXR\,76}).
Though Cha\,H$\alpha$\,7 was also a target of this run,
observations failed for it because of wrong pointing of the telescope 
in service mode.
In programs 75.C-0851(C) and 77.C-0831(A+D),
follow-up spectra of the low-mass star CHXR\,74 were taken,
which was previously revealed as an RV variable source (Joergens 2006a).
Determinations of basic (sub)stellar parameters of the objects can be found in
Comer\'on, Rieke \& Neuh\"auser (1999), Comer\'on, Neuh\"auser \& Kaas (2000), 
and Luhman (2004, 2007).
It is noted that, while the current knowledge of the Cha\,I cloud is complete to 0.01\,$\msun$
(Luhman 2007),
this sample is biased towards brighter BD/VLMS. The reason for this is first, that
the sample was chosen from late-M type objects
that were known at the time of the start of this survey in 2000
(Comer\'on et al. 2000) from shallower surveys, and, second that
three of the faintest objects known at that time, 
\chaha9, \chaha10, and \chaha11, were not included.

\begin{figure}
\centering
\includegraphics[width=0.6\linewidth,clip]{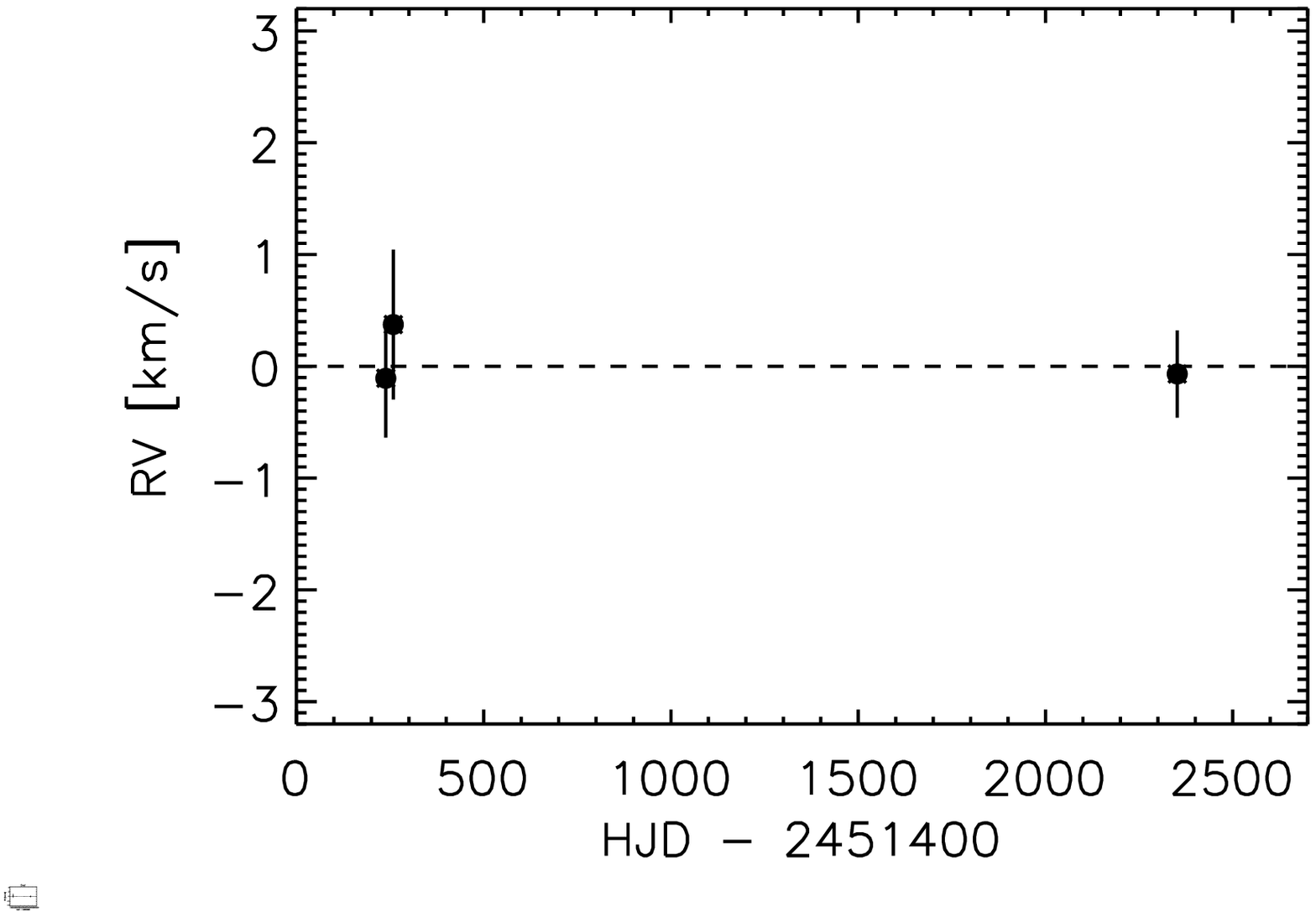}
\includegraphics[width=0.6\linewidth,clip]{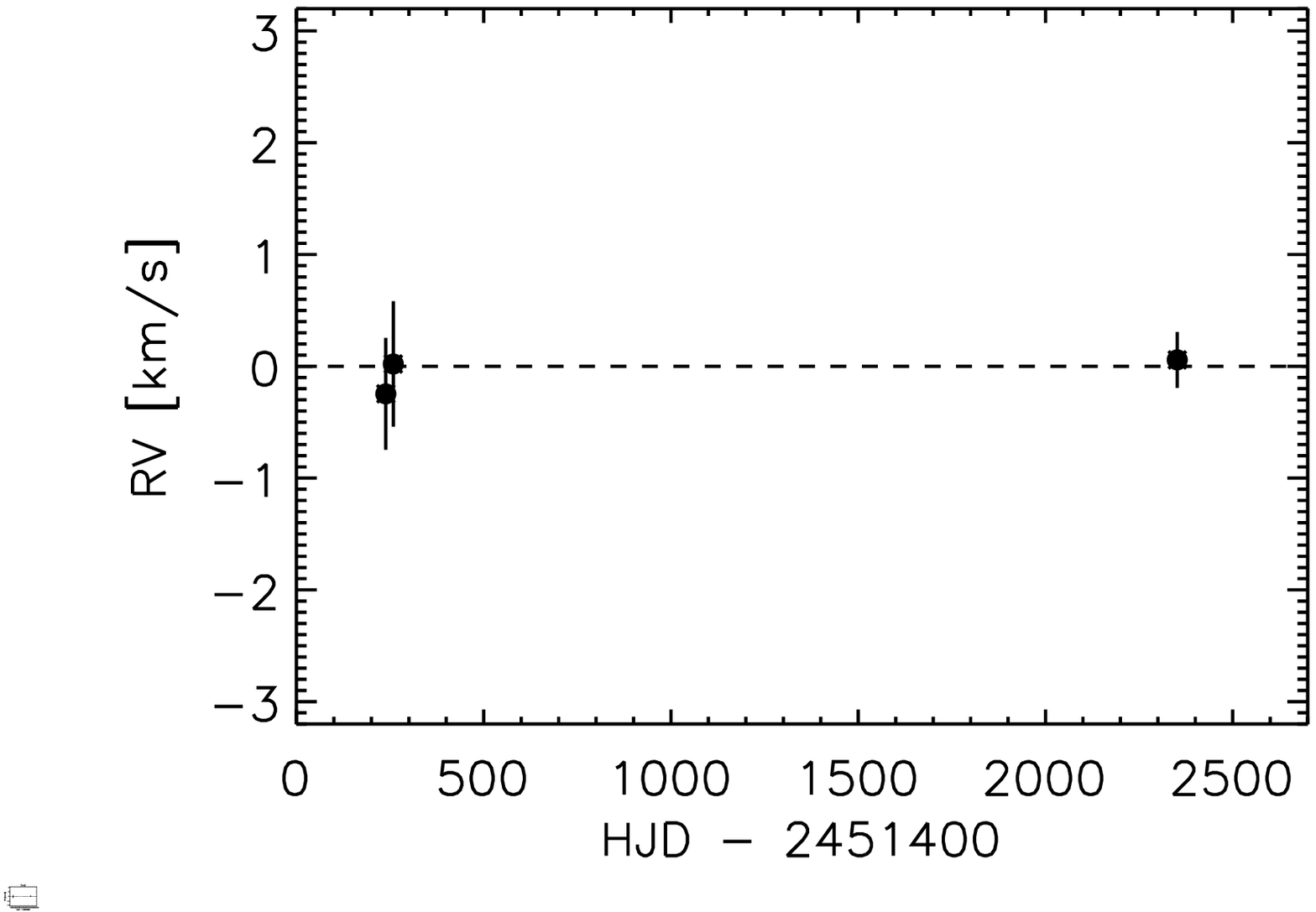}
\includegraphics[width=0.6\linewidth,clip]{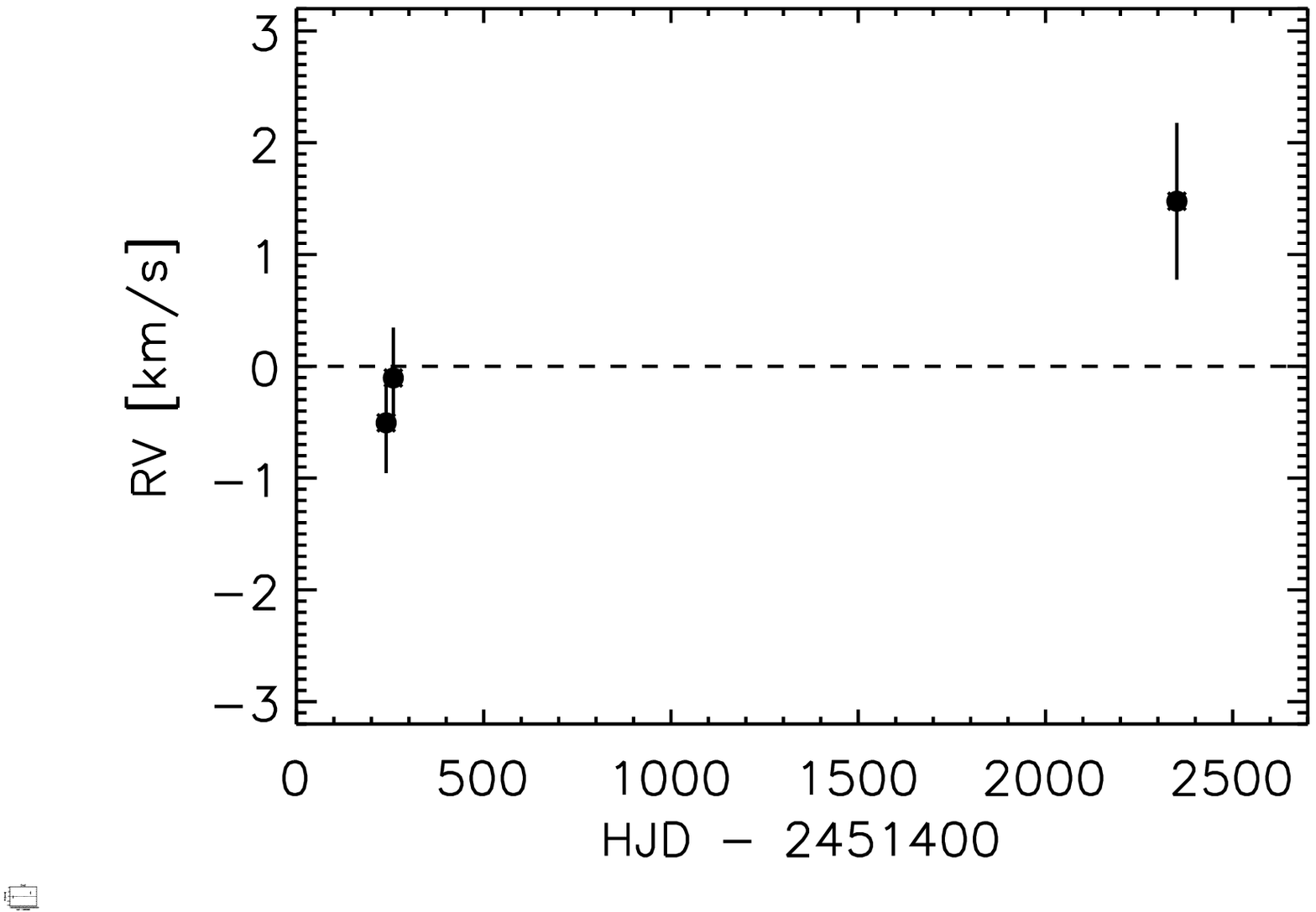}
\caption{
\label{fig:rvs1}
{\bf RV measurements of \chaha1, \chaha2, \chaha3} (top to bottom).
}
\end{figure}

\begin{figure}
\centering
\includegraphics[width=0.6\linewidth,clip]{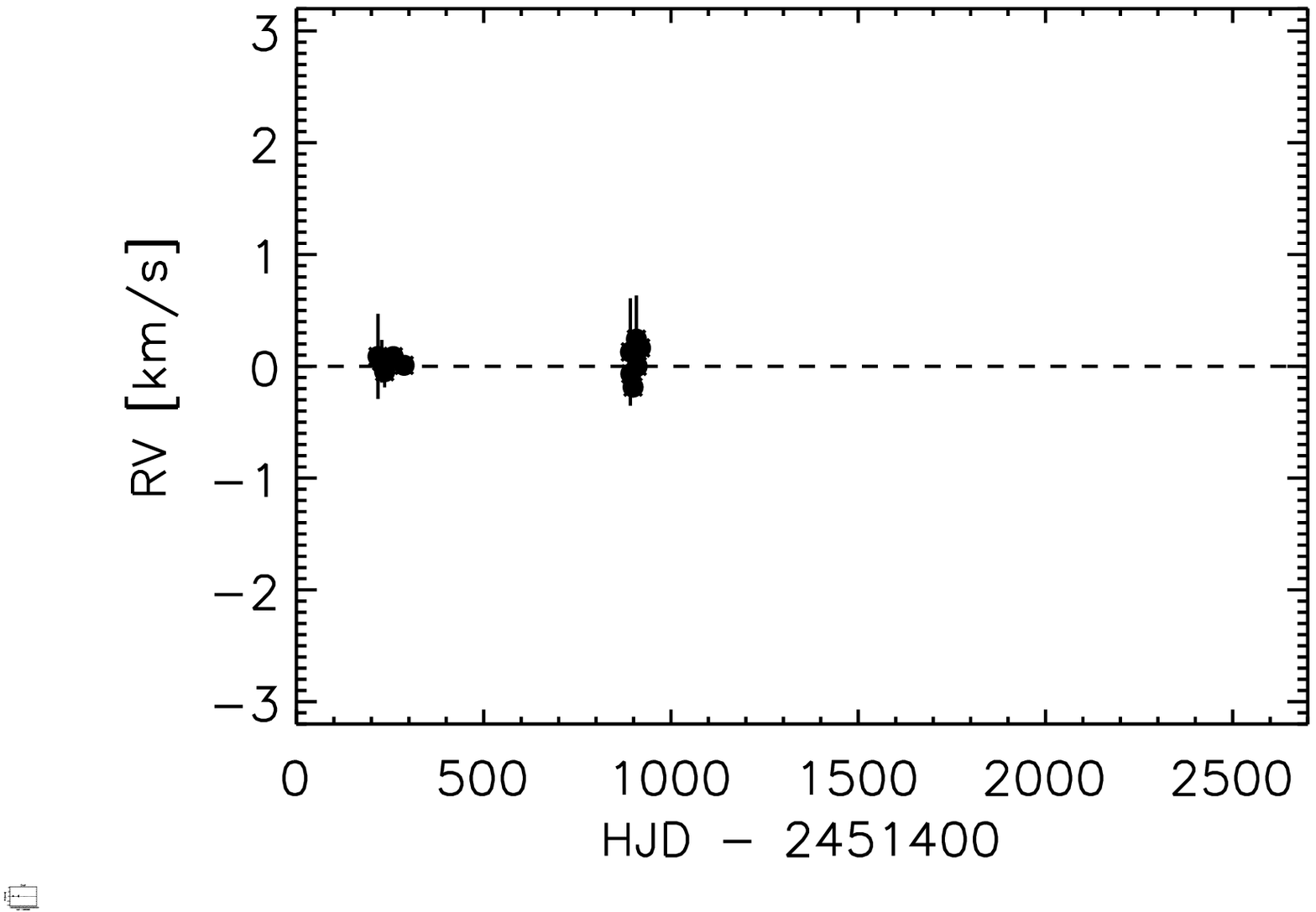}
\includegraphics[width=0.6\linewidth,clip]{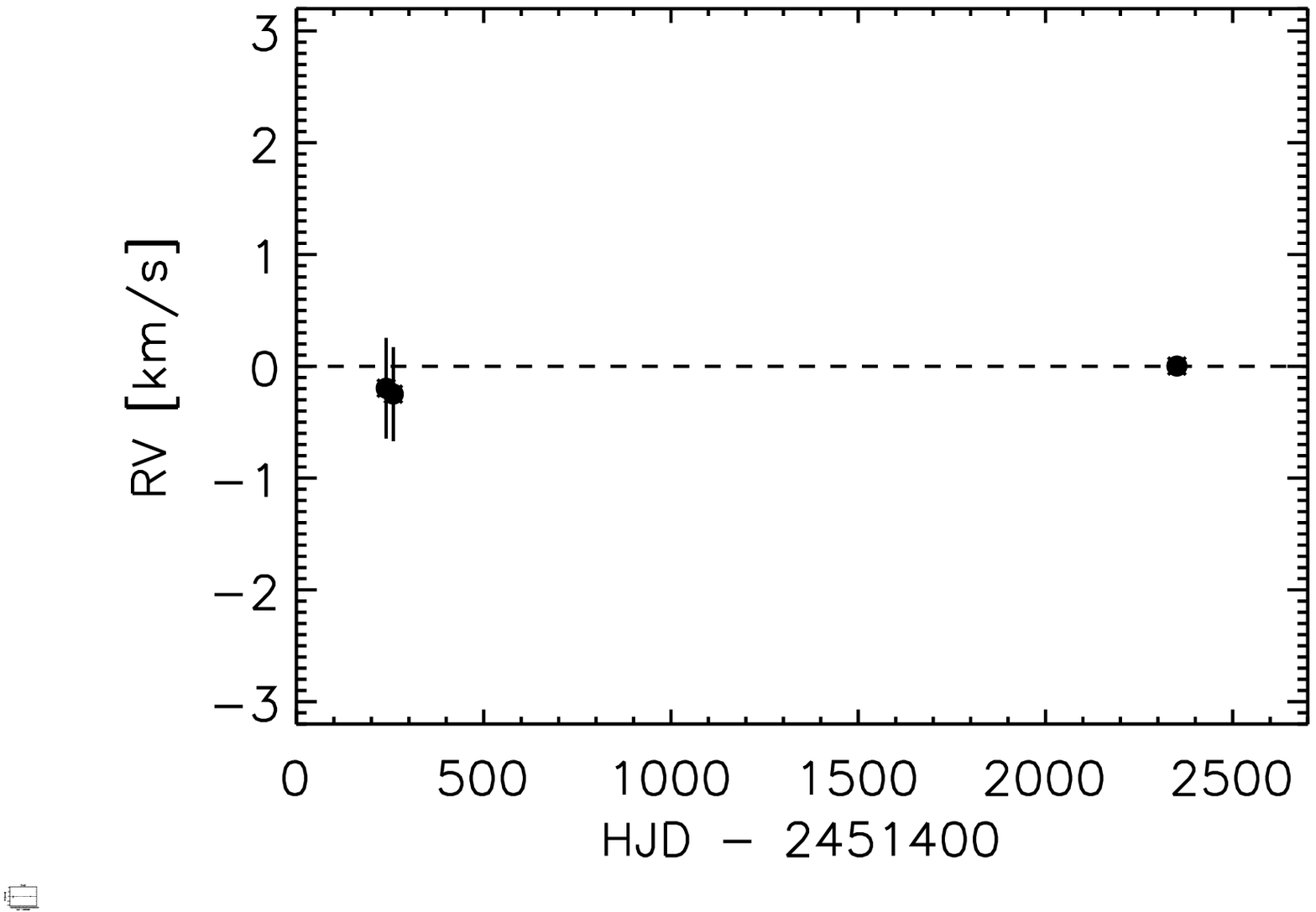}
\includegraphics[width=.6\linewidth,clip]{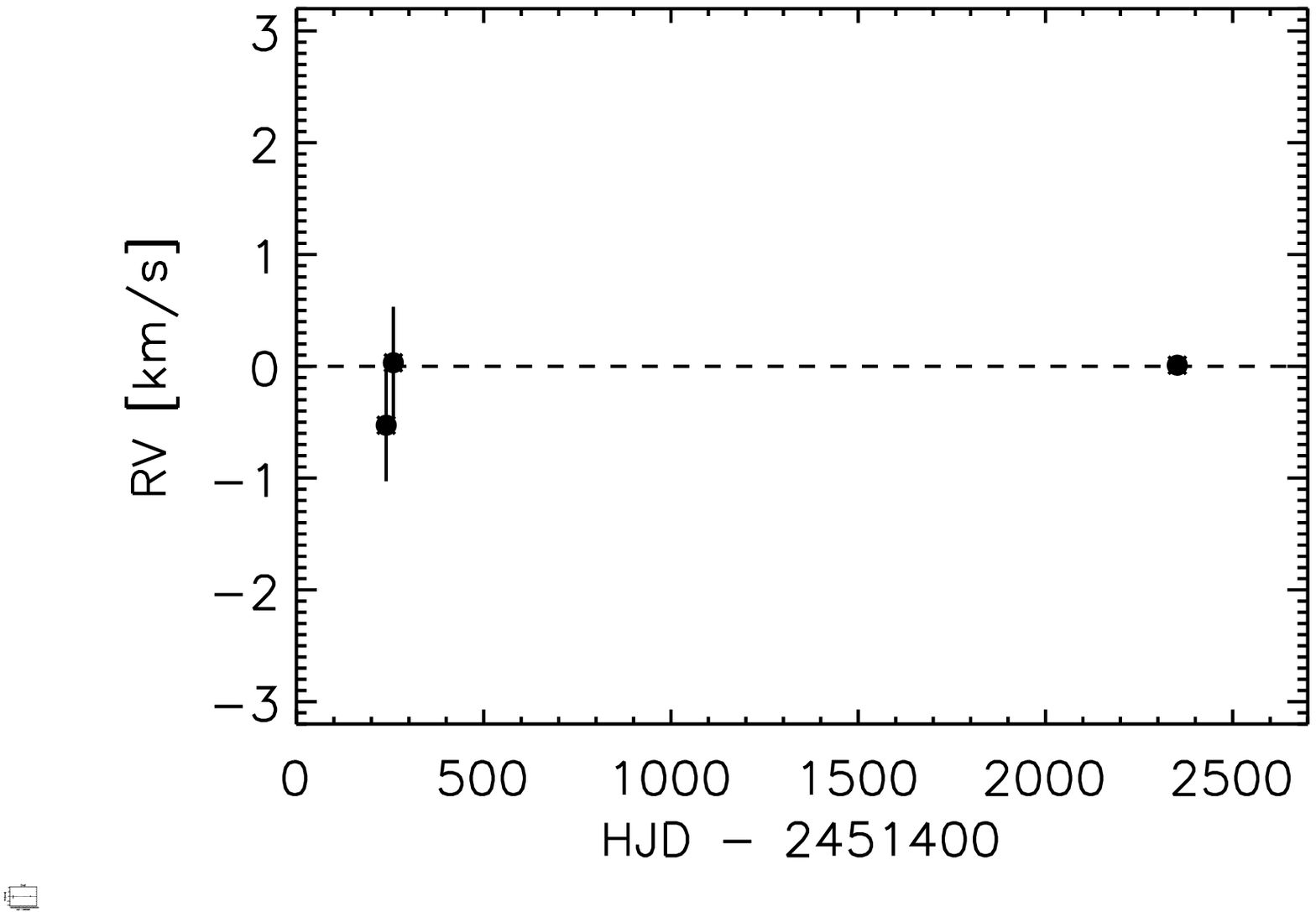}
\caption{
\label{fig:rvs2}
{\bf RV measurements of \chaha4, \chaha5, \chaha6} (top to bottom).
}
\end{figure}

\begin{figure}
\centering
\includegraphics[width=.6\linewidth,clip]{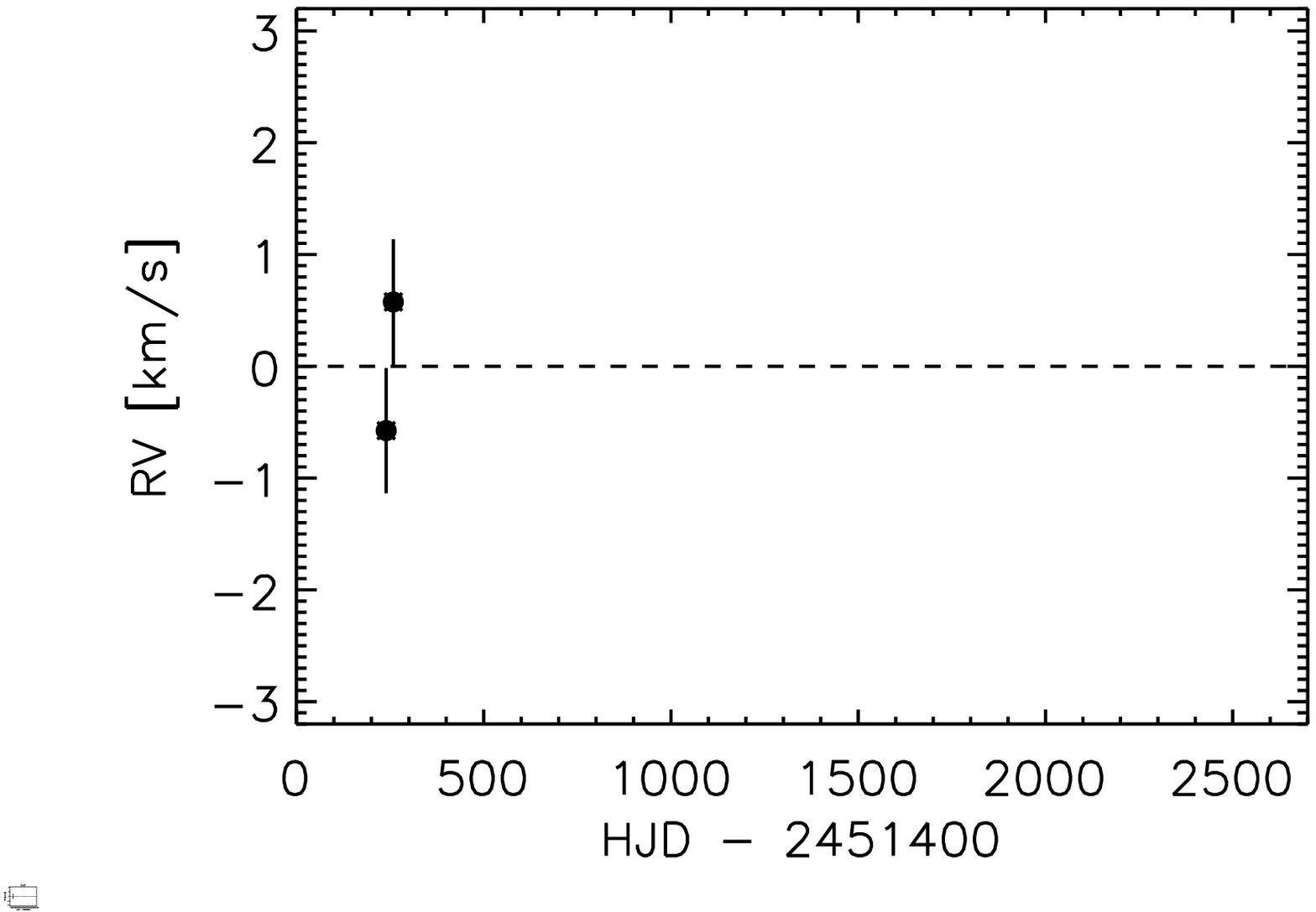}
\includegraphics[width=.6\linewidth,clip]{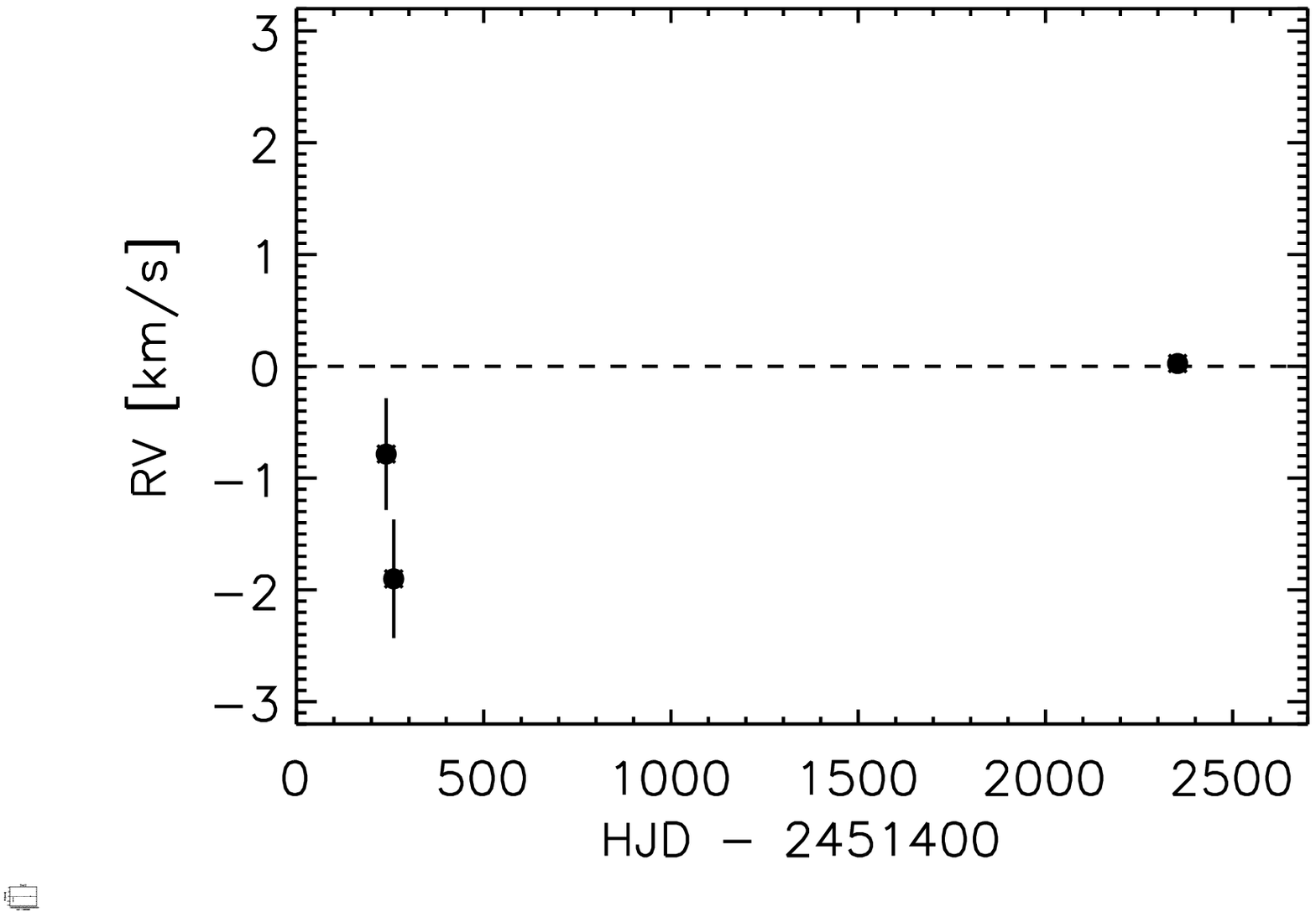}
\caption{
\label{fig:rvs3}
{\bf RV measurements of \chaha7 and \chaha12} (top to bottom).
}
\end{figure}

\begin{figure}
\centering
\includegraphics[width=.6\linewidth,clip]{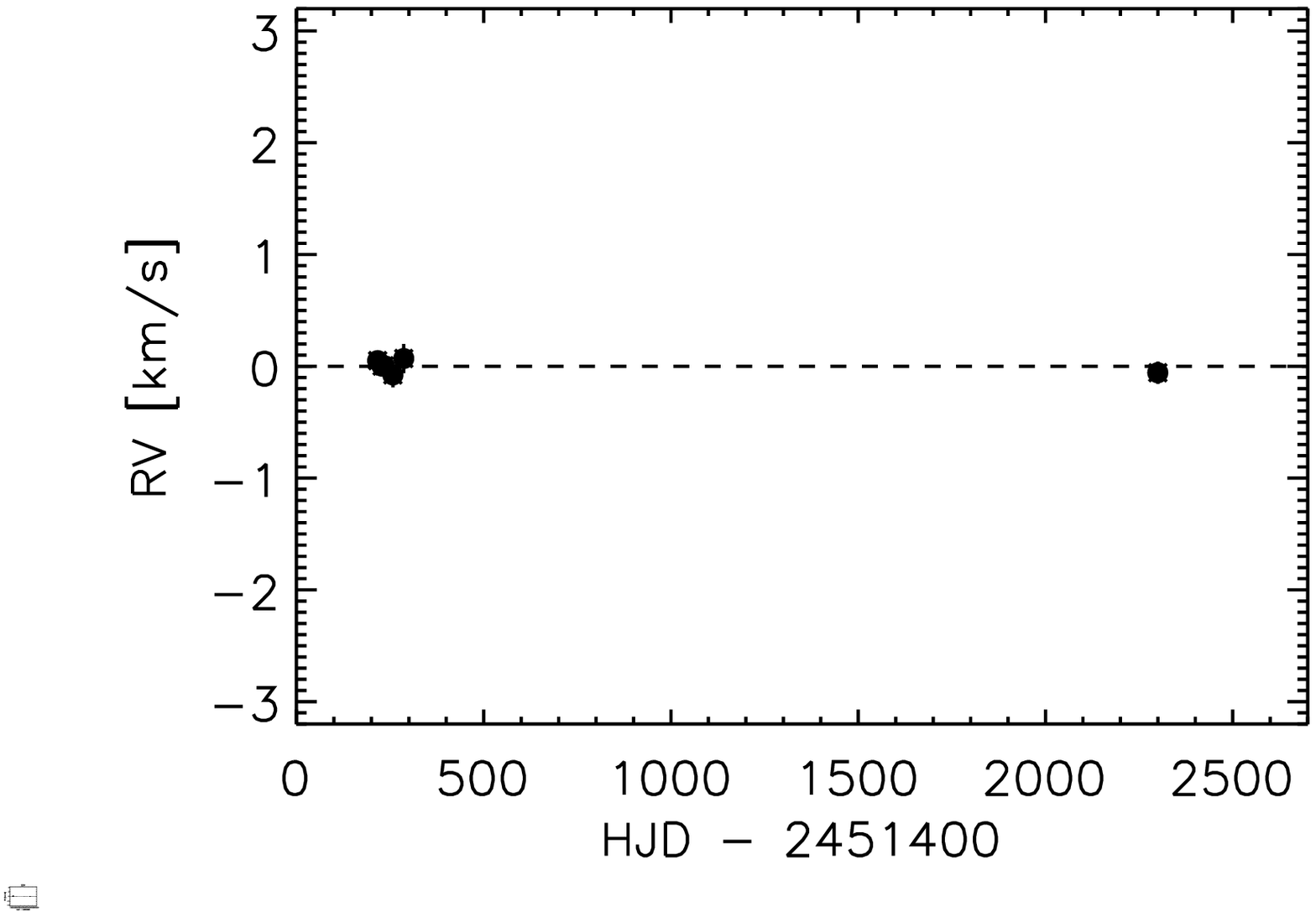}\hfill
\includegraphics[width=.6\linewidth,clip]{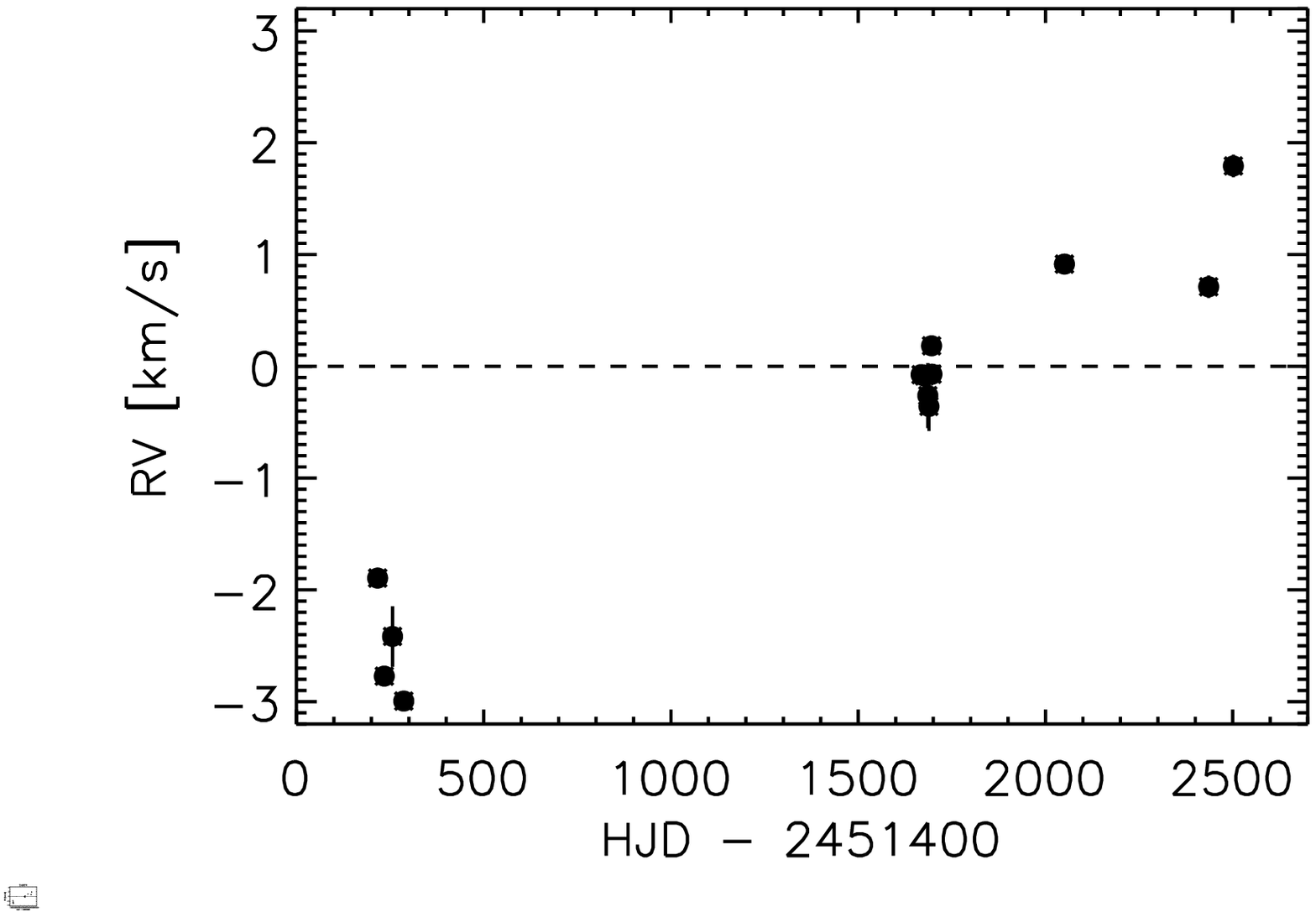}\hfill
\caption{
\label{fig:rvs4}
{\bf RV measurements of B\,34 and CHXR\,74} (top to bottom).
}
\end{figure}

After standard CCD reduction of the data, one-dimensional spectra were extracted and their wavelength
scale established in a first step by means of ThAr lamp exposures.
The RVs were measured from these spectra based on a cross-correlation technique
employing telluric lines for the wavelength calibration. 
Details on the data reduction and analysis can be found in Joergens (2006a).
Each RV determination is based on
two consecutive single spectra that provide two independent measurements of the RV
(an exception here are the 2006 observations of CHXR\,74).
This allows an estimation of the error of the relative 
RVs based on the sample standard deviation for two such data points.
The new RV measurements are summarized in Table\,\ref{tab:rvs}.
These estimated errors of the {\em relative} RVs
range between 30 and 100\,m\,s$^{-1}$ (Table\,\ref{tab:rvs}).
The absolute RVs are subject to an additional error of about 400\,m\,s$^{-1}$
because of the uncertainty of the velocity zero point (cf. Joergens 2006a). 
Figures\,\ref{fig:rvs1}--\ref{fig:rvs4} display the results.
The studied objects have a mean RV of 16.0\,km\,s$^{-1}$
in agreement with membership in the Cha\,I star-forming cloud
(the surrounding molecular gas has a velocity of 15.4\,km\,s$^{-1}$, 
Mizuno et al. 1999).
Their RV dispersion is relatively small: the RV dispersion 
measured in terms of standard deviation is 0.8\,km\,s$^{-1}$ 
and the RVs cover a total range of 2.2\,km\,s$^{-1}$,
consistent with previous kinematic measurements of the objects (Joergens 2006b).
In particular, no outliers are found. 

\section{RV Variability}
\label{sect:RVvari}

There are different approaches to identifying significant variability in
datasets of repeated RV measurements. 
Prato (2007b) classifies an object as RV variable if
the 
root-mean-square (rms) scatter of its repeated 
RV measurements are significantly (five times)
above the average rms scatter of the whole sample,
i.e. assuming that most objects are not intrinsically RV variable.
This method  registers external noise, like RV noise caused by activity, 
by calculating the average rms scatter of the sample. 
Other authors compute the probability with which observed RV values of an object
with individual error estimates
can be described by a constant function in a $\chi^2$ test 
(e.g. Maxted \& Jeffries 2005). 
Objects for which the probability for being RV constant is less than 
a chosen threshold 
are classified as RV variable sources.
The identification of RV variability and, therefore, of spectroscopic binaries
by this method depends crucially on the individual measurement errors.

For the identification of RV variable objects among BD/(V)LMS in Cha\,I,
the second method is applied and 
the $\chi^2$ probability $p$ for each object being RV constant is computed.
The constant function, which is fitted in this test, is given by
the weighted mean of the individual RV values.
The threshold of tolerance for error in the $\chi^2$ test is set to 0.1\% ($p < 10^{-3}$), i.e.
a possible detection of RV variability has a significance of 
at least 3.3\,$\sigma$.
The evaluated data set for BDs and (very) low-mass stars in Cha\,I includes 
the new RV measurements presented here (Table\,\ref{tab:rvs})
as well as previous ones by Joergens (2006a) and Joergens \& M\"uller (2007).
Table\,\ref{tab:chi2prob} summarizes the relevant characteristics of these data
for each studied object 
(half peak-to-peak RV differences, the weighted mean RV, 
the error of this mean, which takes into account an uncertainty of 400\,m\,s$^{-1}$ for the
zero point of the velocity,
the number of observations, the minimum time separation
between two RV measurements, and the covered time base).
For the variability classification, in addition to the listed internal RV errors
(in most cases standard deviation of two consecutive measurements)
a minimum error of 0.3\,km\,s$^{-1}$ has been applied to account for 
fluctuations in the standard deviation for a small number of measurements
and/or for systematic errors.
This minimum error is derived by an estimate of the systematic errors based on 
the night-to-night rms scatter of all BD/VLMS in the sample of 0.32\,km\,s$^{-1}$ on average
(i.e. without making any assumptions on the internal errors or
the variability status), which is 
in good agreement with their average internal error of 0.29\,km\,s$^{-1}$.
The calculated $\chi^2$ probabilities are listed in Table\,\ref{tab:chi2prob}. 
Both \chaha8 and CHXR\,74 have values below
the chosen threshold ($p < 10^{-3}$), hence, are identified as RV variables.
For both, \chaha8, which was recently shown to have a BD companion
(Joergens 2006a; Joergens \& M\"uller 2007) and
for CHXR\,74, for which a longterm RV trend was observed (Joergens 2006a), 
these are confirmations of previous findings of spectroscopic companions.
For CHXR\,74, there are two epochs of RV data added in this work
providing further constraints on the orbit of the companion. 
See Sect.\,\ref{sect:chxr74} for more details on CHXR\,74.

\begin{figure}
\centering
\includegraphics[height=\linewidth,angle=90]{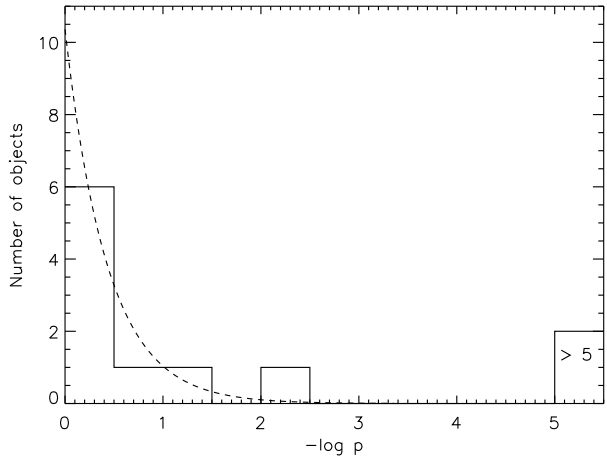}
\caption{
\label{fig:prob}
{\bf Histogram of the $\chi^2$ probability $p$} for fitting the observed
(relative) RV values of the BD/(V)LMS in Cha\,I
studied here with a constant function. 
Objects with $p < 10^{-5}$ are shown in the right-most bin,
\chaha8 and CHXR74.
Based on the threshold $p < 10^{-3}$, they are classified as binaries.
Overplotted is the expected distribution for nonvariable
objects given the estimated uncertainties. It shows that the error
estimates are reliable.
}
\end{figure}

Figure\,\ref{fig:prob} shows the distribution of $\chi^2$ probabilities.
For the RV constant objects, it resembles 
the expected distribution of $-\log p$ for random fluctuations
alone normalized to the number of constant objects (dashed line).
This shows that the applied error estimates are reliable
and, in particular, that they have not been underestimated.
The RV variable objects, \chaha8 and CHXR\,74, appear in the rightmost
bin of the histogram in Fig.\,\ref{fig:prob}. They are clearly separated 
from the nonvariable group.

The performed variability classification by a $\chi^2$ test is based on 
the assumption that the errors are independent of the data.
Applying a minimum error given by the average scatter
of the measurements, as done here, accounts for non-Gaussian errors and
partly ensures such an independence.
To further increase the confidence, the result has been
double checked by means of an analysis of variance (ANOVA) test,
which is a generalization of the student's t-test to more than two measurements and
which is independent of error estimates.
The F probabilities were calculated
using the individual RV measurements, i.e. for RV values that are
based on two consecutive RV measurements, the two individual RV values were used.
It is found that the outcome of this analysis of variance is basically consistent 
with that of the $\chi^2$ test; in particular, the same objects were found 
as significantly variable sources. 

One can use the presented data to assess the level of RV noise 
for very young BD/VLMS (in Cha\,I) by quantifying the 
night-to-night rms scatter of only the constant BD/VLMS in the sample, which
is 0.31\,km\,s$^{-1}$ on average. 

\begin{figure}
\centering
\includegraphics[height=\linewidth,angle=90]{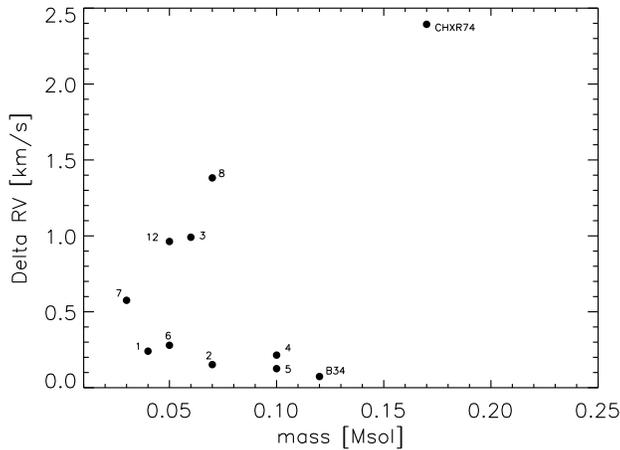}
\caption{
\label{fig:deltaRV}
{\bf RV differences vs. object mass.} 
Plotted are half peak-to-peak differences in the observed RVs.
Each data point is labeled with the corresponding object name, the numbers
denote the Cha\,H$\alpha$ objects.
The RV variable objects (CHXR\,74, \chaha8) have 
$\Delta$RV $\gtrsim$ 1.4\,km\,s$^{-1}$, while the majority of RV constant
objects have $\Delta$RV $\leq$0.3--0.6\,km\,s$^{-1}$, with the exception
of \chaha3 and \chaha12, which have $\Delta$RV $\sim$1\,km\,s$^{-1}$.
}
\end{figure}

Another way to look at the results of this survey is to
consider the recorded RV differences. 
Figure\,\ref{fig:deltaRV} displays 
half peak-to-peak RV differences $\Delta$RV versus the object mass,
as adopted from Comer\'on et al. (1999, 2000)\footnote{Luhman (2004, 2007) derives 
systematically higher effective temperatures and luminosities
for the objects leading to higher mass estimates 
when compared with evolutionary tracks,
e.g. 0.1\,$\msun$ for \chaha8 instead of 0.07\,$\msun$.}:
Objects classified as RV constant have $\Delta$RV $\leq$1\,km\,s$^{-1}$,
and the majority of them have even smaller 
$\Delta$RV below 0.3\,km\,s$^{-1}$. 
The two RV variables, on the other hand, exhibit 
RV differences $\Delta$RV above 1.4\,km\,s$^{-1}$.
Though clearly separated from the vast majority of the RV constant objects
in this diagram, it is remarkable that RV variability
occurs with such relatively small amplitudes. 
This is discussed further in Sect.\,\ref{sect:disc}.

It is noted that, because of the small variability amplitudes, 
none of the objects would have been identified as RV variable source
when applying the criterion of Prato (2007b), namely to require that the rms scatter
has to be five times greater than the average rms scatter of the whole sample
to classify an object as RV variable:
The rms scatter of the RV variable objects, 
\chaha8 and CHXR\,74, is 1.0 and 1.5\,km\,s$^{-1}$, 
respectively, while the average rms scatter of the whole sample is 0.6\,km\,s$^{-1}$  
and that of the subsample of BD/VLMS is 0.5\,km\,s$^{-1}$.
It is further noted that, if the
intrinsic errors of the RV measurements would have been as large as in the
survey of Basri \& Reiners (2006, 1.4\,km\,s$^{-1}$ on average), also 
none of the objects would have been identified as RV variable.

\begin{figure}[t]
\centering
\includegraphics[height=\linewidth,angle=90]{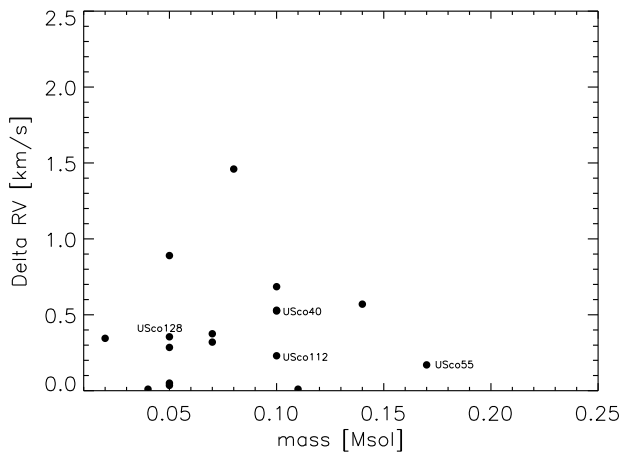}
\caption{
\label{fig:deltaRV_K}
{\bf RV differences for BD/VLMS in USco and $\rho$\,Oph} 
based on data from Kurosawa et al. (2006).
$\Delta$RV as in Fig.\,\ref{fig:deltaRV}. 
Objects classified as RV variables by these authors are denoted with their names.
The two data points with the largest RV difference correspond 
to GY310 and USco101, which are classified as RV constant.
}
\end{figure}

In a similar survey but with more restricted separation range,
Kurosawa, Harries \& Littlefair (2006) find four spectroscopic BD/VLM binary candidates
in USco and $\rho$\,Oph. The recorded half peak-to-peak RV differences for them
($\Delta$RV = 0.2--0.5\,km\,s$^{-1}$) are significantly less 
than for the variable BD/VLMS in Cha\,I (1--2\,km\,s$^{-1}$).
Figure\,\ref{fig:deltaRV_K} shows $\Delta$RV plotted versus mass for
the data from Kurosawa et al. (2006). It can be seen here
that the four RV variable objects 
(marked in the figure with their names)
cannot be distinguished from the RV constant 
objects based on larger RV differences.
(The average intrinsic error of the whole sample 
is 0.42\,km\,s$^{-1}$, five objects in the sample
have RV errors $\leq$0.15\,km\,s$^{-1}$, four of them are found to be significantly
variable.)
While it has to be further investigated whether it can be excluded that the detected variability is caused 
by systematic errors, 
if we interpret them as RV semi-amplitudes caused by companions, these companions
would have masses M$_2 \sin i$ in the planetary regime (1-2 Jupiter masses) 
assuming a separation of $\leq$0.1\,AU.

\section{Binary fraction and probed orbital separations}
\label{sect:binfrac}

Based on the statistical analysis in the previous section, 
two objects (\chaha8, CHXR74) among a sample of eleven BDs and 
(very) low-mass stars (spectral type M4.25 and later;
M$\lesssim 0.25\,\msun$) are found with significant RV variations 
and are, therefore, considered as spectroscopic binary systems.
The uncertainty of the measured binary fraction is estimated
by searching for the fraction of binaries at which the
binomial distribution P$_B$(x,n,p) for x positive events in n trials
falls off to 1/$e$ of its maximum value (cf. Basri \& Reiners 2006).
The observed binary fraction of the whole sample is 18$^{+20}_{-12}$\,\% (2/11).
When considering only the BD/VLMS in the sample, i.e. the ten objects with
M$\lesssim 0.1\,\msun$, we observe one RV variable (\chaha8), i.e. 
a BD/VLM binary fraction of 10$^{+18}_{-8}$\,\% (1/10).

The orbital separation range that can be probed by an RV survey 
for companions depends mainly on the timely sampling of the measurements
and on their uncertainties. 
To assess the detection efficiency as a function of 
the orbital separation of this RV survey, 
a Monte-Carlo simulation of RV measurements for 
randomly selected binaries from a model was performed
using the time spacing and error estimates as given by the real observations
and again applying a minimum error of 0.3\,km\,s$^{-1}$.
For each simulated binary, it is then checked whether it would have been detected
as RV variable in this survey, i.e. if its probability for being 
RV constant is lower than the chosen threshold of $p < 10^{-3}$ (cf. Sect.\,\ref{sect:RVvari}).
A similar but more sophisticated simulation has been performed by Maxted \& Jeffries (2005)
and applied in a simplified version by other authors (e.g.
Kurosawa et al. 2006).  
The procedure is described in these works and 
the following summarizes only the basic steps and assumptions applied here.
For each object of the sample, the relative RV errors, number of RV data points,
and their temporal separation as given by the real observations are considered.
Furthermore, the primary mass estimates from Comer\'on et al. (1999, 2000) are used.
For each semi-major axis in the range between 0.006 and 10\,AU, 
10$^5$ binaries are simulated by (i) assuming a flat mass ratio distribution 
between $q\equiv M_2/M_1=$0.2 and 1.0, i.e.
selecting the mass ratio randomly from this range; 
(ii) assuming circular orbits, i.e. setting the eccentricity to zero and not considering
the longitude of periastron;
(iii) selecting the orbital inclination $i$ randomly from a uniform $\cos i$ distribution;
and by
(iv) selecting the orbital phase for the
first simulated RV randomly between 0 and 1.
These assumptions are justified by the results of Maxted \& Jeffries (2005), who find
that the detection probability is mainly insensitive to the
distribution of mass ratio and eccentricity.
The minimum considered separation in this simulation is 0.006\,AU. This allows
a distance between central object and companion of more than
2.4 times the radius (Roche criterion for stable orbits) and assumes a size of half a solar radius.
(Very young BDs have relatively large radii, 
e.g. the radius of a BD in Orion with mass 0.034\,$\msun$ was measured
to 0.51\,$\rsun$=0.0024\,AU, Stassun et al. 2006.)
For each binary selected in the described way, RV measurements are simulated by 
calculating the RV for the first orbital phase, evolving the orbit 
based on the given time sequence of the real observations, and  
calculating the subsequent RV measurements.
These simulated RV data, along with the RV errors of the real observations,
are then subject to the same RV variability check as performed for the observations. 
Finally, the detection probability is given by
the number of detected
binaries compared to the total number of 10$^5$ simulated binaries for each semi-major axis.
It is noted that line-blending is not taken into account in this simulation.
This effect is most relevant for nearly equal-mass systems.
Maxted et al. (2008) show 
that neglecting line-blending can lead to an overestimate of the
detection efficiency of up to 15\%. This has to be kept in mind when interpreting the 
detection probabilities calculated here and in the framework of 
other RV surveys of BD/VLMS that
neglect this effect (e.g. Kurosawa et al. 2006; Basri \& Reiners 2006).

\begin{figure*}
\centering
\includegraphics[width=\linewidth,angle=0]{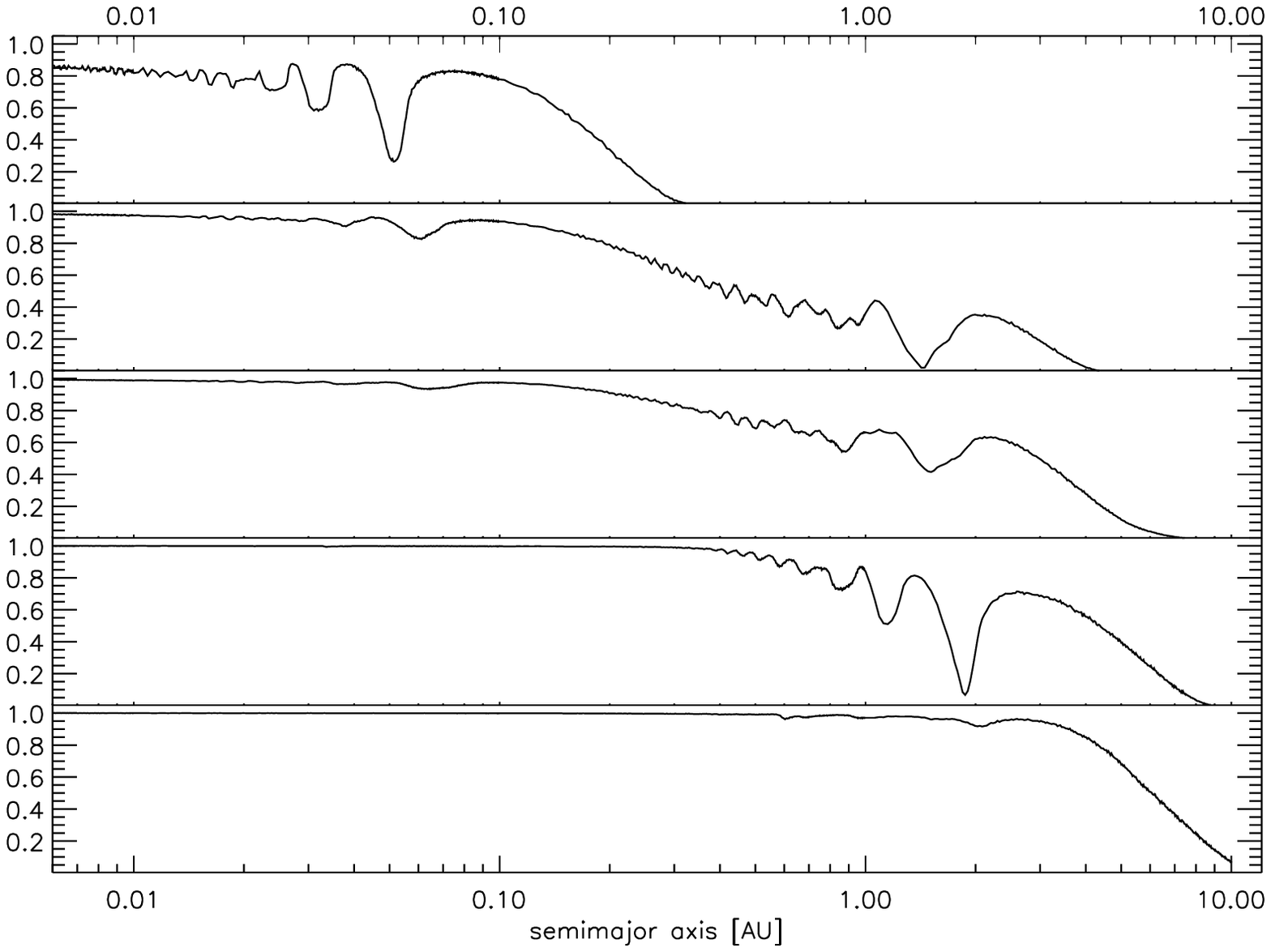}
\caption{
\label{fig:detectprob}
{\bf Detection probability as function of semi-major axis.} 
Based on a Monte Carlo simulation of RV measurements 
for randomly selected binaries using the measurement errors
and time separations of the real observations.
From top to bottom: \chaha7; 
average of \chaha1 and \chaha3; 
average of \chaha2, \chaha4, \chaha5, \chaha6, \chaha8, and \chaha12; 
B\,34; and CHXR\,74.
As indicated, some curves are the average detection probability 
for two or more objects because observations have similar quality for those.
}
\end{figure*}

Figure\,\ref{fig:detectprob} shows the results of the Monte-Carlo simulation
for all objects in the order of increasing 
sensitivity for larger separations. 
The majority of objects in the sample 
(\chaha1, \chaha2, \chaha3, \chaha5, \chaha6, \chaha12)
was observed with a temporal spacing of
about 20 days and re-observation after more than about 2100 days (5.8 years). 
Four objects (\chaha4, \chaha8, B\,34, CHXR\,74) 
were observed with a higher cadence. For them, between 6 and 13 independent
RV data points were obtained covering a total time base between 700 and 2540 days (2-7 years).
As can be seen from Fig.\,\ref{fig:detectprob},
the range of orbital separations that can be probed with a 50\% or higher
detection probability is 
$\lesssim$1\,AU for \chaha1 and \chaha3, 
$\lesssim$ 3\,AU for \chaha2, \chaha4, \chaha5, \chaha6, \chaha8, and \chaha12, 
$\lesssim$4\,AU for B\,34, and
$\lesssim$6\,AU for CHXR\,74, respectively.
For \chaha7 (Fig.\,\ref{fig:detectprob}, top panel), 
the probed separation range is 
significantly narrower ($\lesssim$0.2\,AU) than for the other 
objects because scheduled re-observations failed 
(cf. Sect.\,\ref{sect:rvs}). For this BD, the total time base of its 
RV measurements is 20 days.
On average, the semi-major axis range of $\leq$3\,AU was probed with a 
50\% or higher detection rate.

The detection probability can be translated into a rate of binary systems present in the 
sample but missed in the survey. 
However, the absolute number of missed binaries in this survey cannot be quantified
because the orbital period distribution is unknown for BD/VLMS. 
Therefore, no corrected binary fraction is derived
based on the observed binary fraction.

The maximum separation to which an RV survey is sensitive depends
not only on the total covered time but, 
in addition, also crucially on the measurement errors.
For example, \chaha5 has been observed
with almost exactly the same time pattern as
e.g. \chaha1 and \chaha3; however, 
the separation range probed with 50\% detection efficiency
is $\lesssim$3\,AU for \chaha5 compared 
to $\lesssim$1\,AU for \chaha1 and \chaha3, respectively.
The only difference here is the RV error estimates, which are on average
0.4\,km\,s$^{-1}$  
for \chaha5 and 0.5\,km\,s$^{-1}$ for the other two objects
(applying a minimum error of 0.3\,km\,s$^{-1}$), respectively. 

\begin{figure*}
\centering
\includegraphics[height=\linewidth,angle=90]{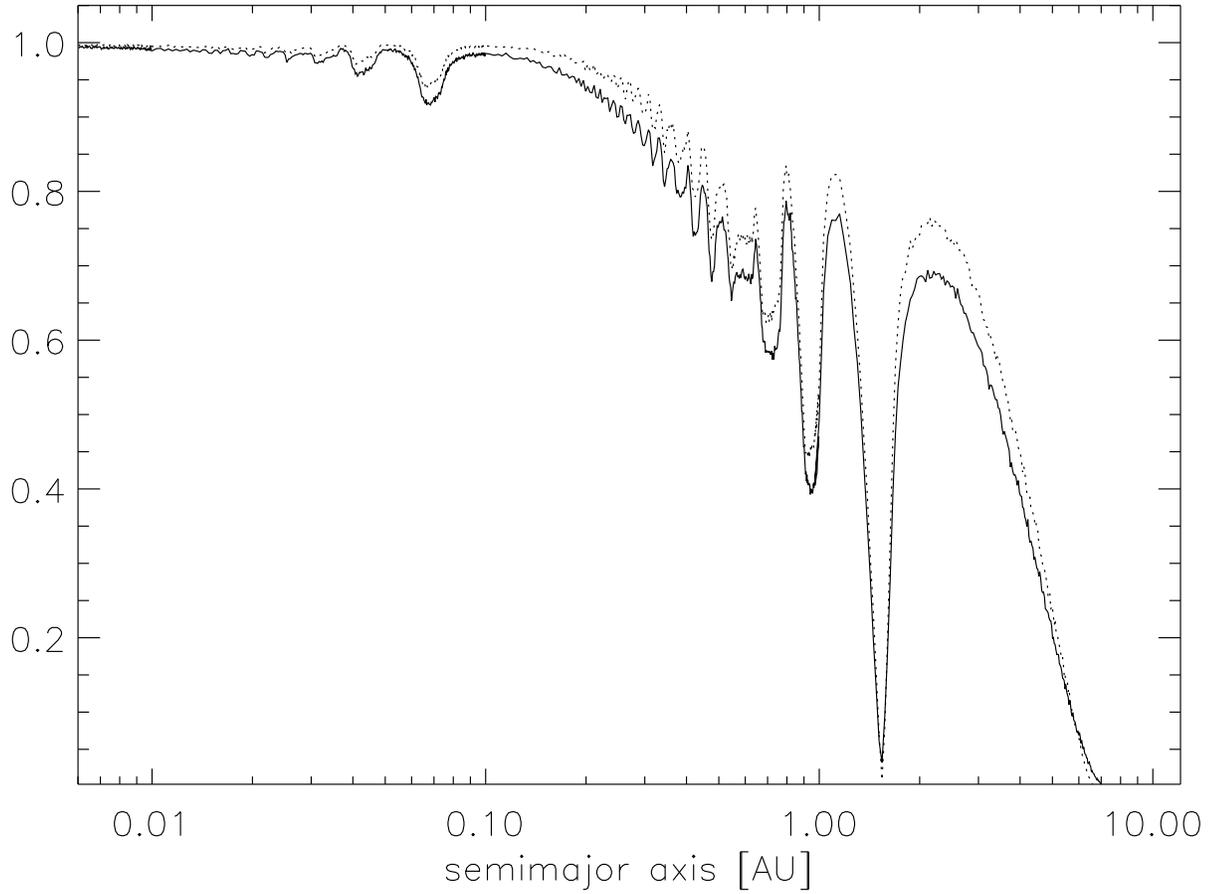}
\caption{
\label{fig:detectprob_comp}
{\bf Dependence of detection probability on 
assumption about inclination.} 
Both curves are 
calculated for $M_1$=0.07\,$\msun$, $\sigma_{RV}$=0.29\,km\,s$^{-1}$, 
$t_{obs}$=[0,20\,d,2000\,d], and $q$ selected randomly from [0.2,1].
Assuming a fixed mean inclination value (dotted line), as in Basri \& Reiners (2006),
slightly overestimates the detection efficiency compared to the
realistic case of random orientation (solid line).
For the fixed inclination case,
the mean value of the inclination distribution of 57.3$^{\circ}$ is used.
}
\end{figure*}

\begin{figure*}
\centering
\includegraphics[height=\linewidth,angle=90]{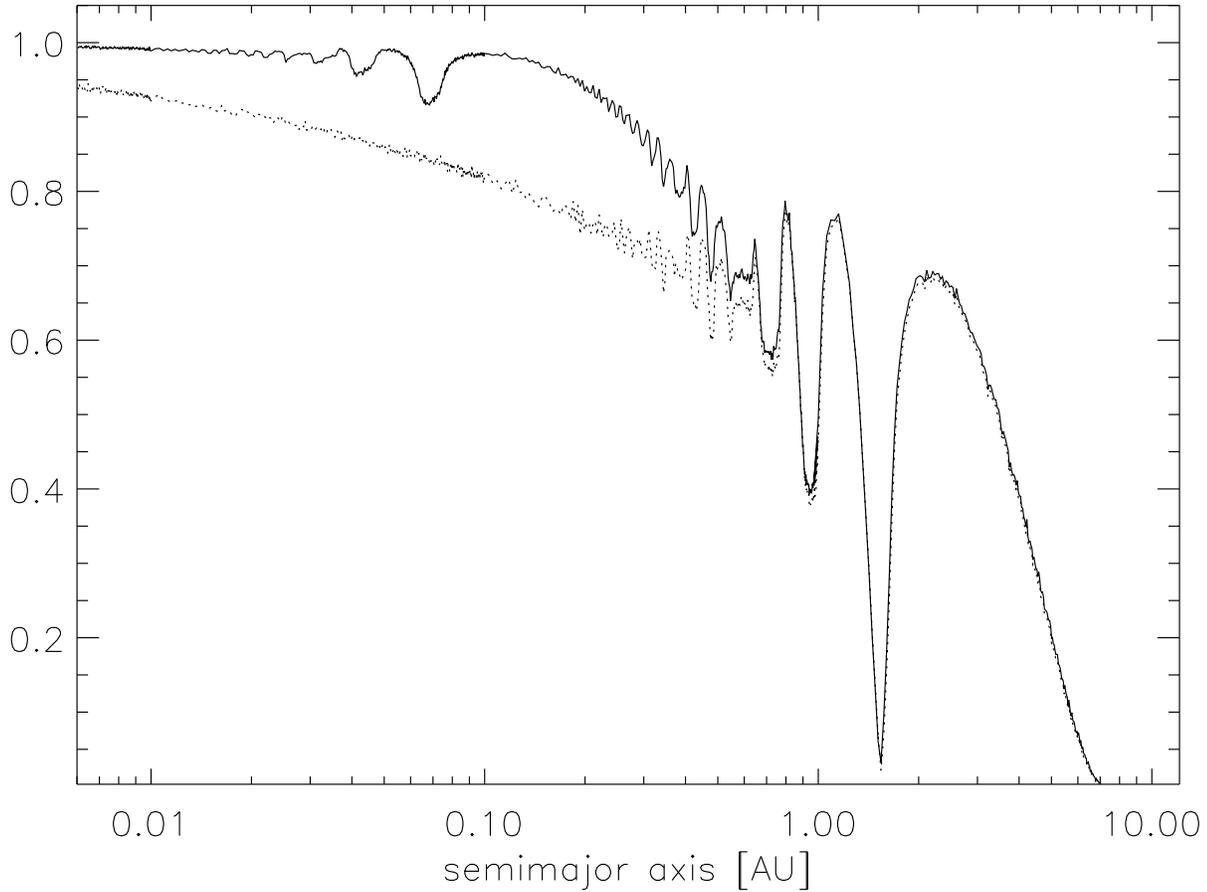}
\caption{
\label{fig:detectprob_comp2}
{\bf Dependence of detection probability on 
number of observations.} 
Both curves are 
calculated for $M_1$=0.07\,$\msun$, $\sigma_{RV}$=0.29\,km\,s$^{-1}$, 
a maximum time span of 2000\,d,
$q$ selected randomly from [0.2,1], and random orientation.
An observing schedule based on three observations with
$t_{obs}$=[0,20\,d,2000\,d] (solid line) provides a superior 
detection efficiency compared 
to the case
of only two observations separated by 2000\,d ($t_{obs}$=[0,2000\,d], dotted line).
Thus, third-epoch observations between the observations with the largest
time difference can have a significant effect on the probability distribution.
}
\end{figure*}

Figure\,\ref{fig:detectprob_comp} shows that 
assuming a fixed value for the orbital inclination
in the Monte-Carlo simulation,
as applied by Basri \& Reiners (2006), leads to a slight overestimation
of the detection efficiency for larger orbital separations
compared to the realistic case of random orientation.
In Fig.\,\ref{fig:detectprob_comp2}, it is demonstrated that
third-epoch observations between the observations with the largest
time difference can have a significant effect on the probability distribution.
While such intermediate third-epoch observations do not
influence the maximum probed separation, they can increase the detection rate
at smaller separations, here by up to 20\%.

\begin{figure*}
\centering
\includegraphics[width=\linewidth,angle=0]{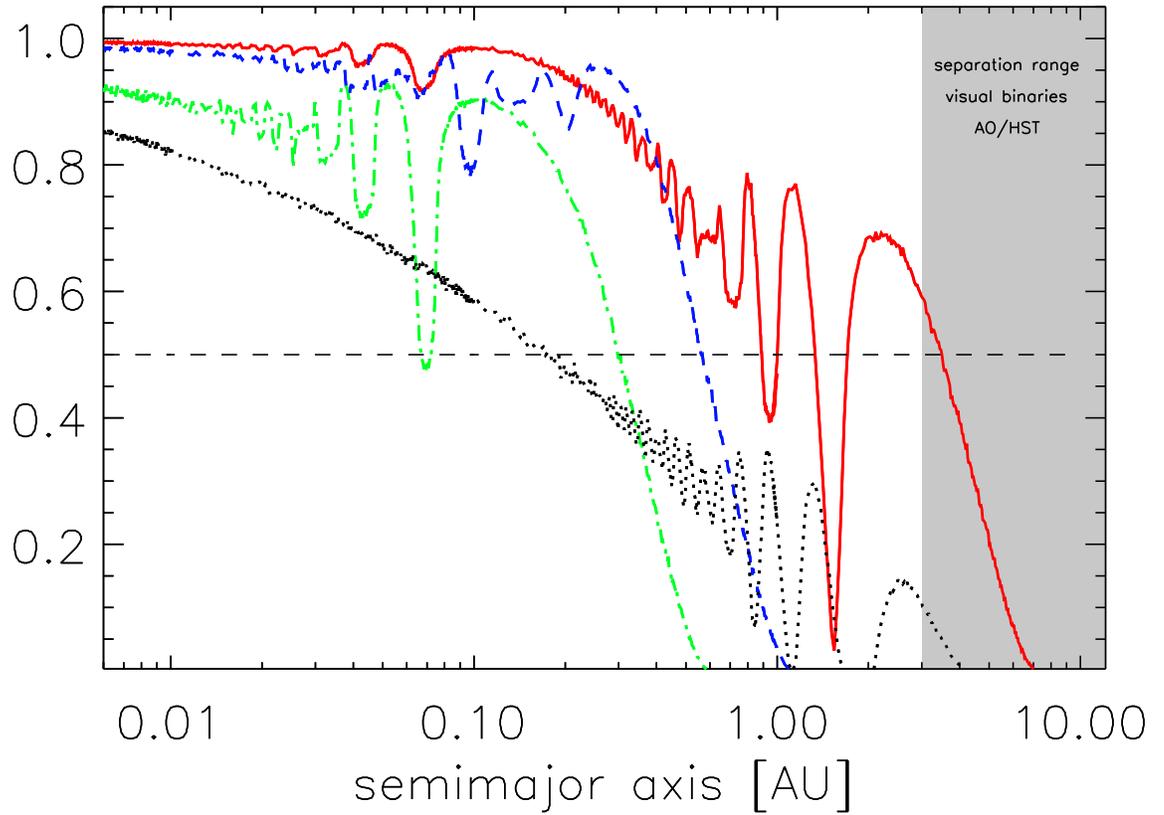}
\caption{
\label{fig:vgl_surveys1}
{\bf Comparison of detection probability of different BD/VLMS RV surveys.} 
All curves were calculated for $q$ selected randomly from [0.2,1], random orientation, 
variability threshold of $p < 10^{-3}$.
Solid (red) line: this work (calculated for $M_1$=0.07\,$\msun$, $\sigma_{RV}$=0.29\,km\,s$^{-1}$, 
$t_{obs}$=[0,20\,d,2000\,d]);
dotted (black) line: Basri \& Reiners 2006 ($M_1$=0.15\,$\msun$, $\sigma_{RV}$=1.3\,km\,s$^{-1}$, 
$t_{obs}$=[0,1650\,d]); 
dashed (blue) line: Guenther \& Wuchterl 2003 ($M_1$=0.1\,$\msun$, $\sigma_{RV}$=0.64\,km\,s$^{-1}$, 
$t_{obs}$=[0,31\,d,85\,d]); 
dash-dotted (green) line: Kurosawa et al. 2006 ($M_1$=0.07\,$\msun$, $\sigma_{RV}$=0.42\,km\,s$^{-1}$, 
$t_{obs}$=[0,20\,d]).
See online version for a color figure.
}
\end{figure*}

\begin{figure*}
\centering
\includegraphics[width=\linewidth,angle=0]{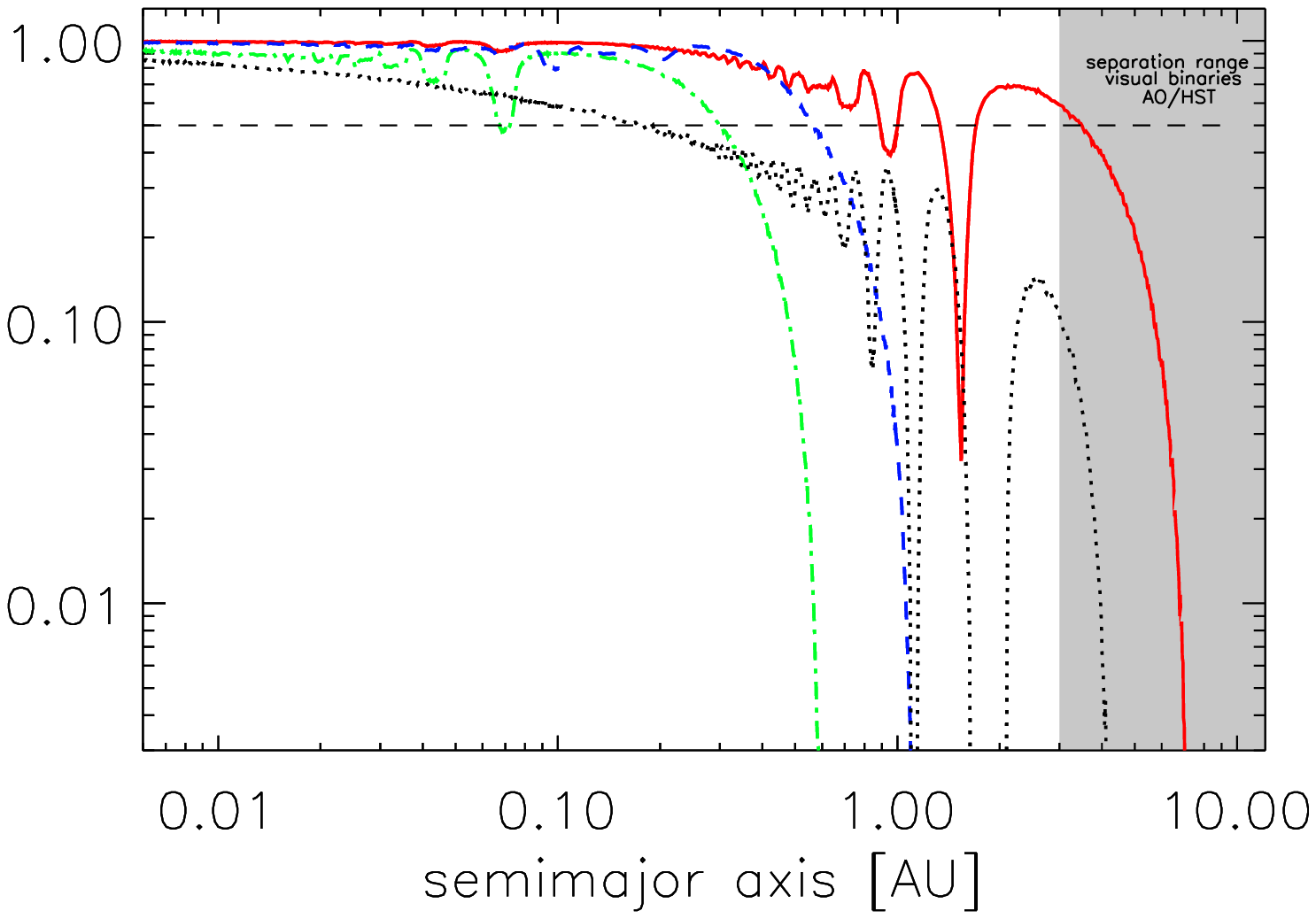}
\caption{
\label{fig:vgl_surveys2}
{\bf Same as Fig.\,\ref{fig:vgl_surveys1} but with logarithmic y-axis
to facilitate comparison with other publications.
} 
See online version for a color figure.
}
\end{figure*}

In the following, the detection efficiency vs. orbital separation
of several current RV surveys of BD/VLMS 
(Guenther \& Wuchterl 2003; Kurosawa et al. 2006; 
Basri \& Reiners 2006; this work)
is compared in a uniform way.
For this purpose, the detection probability is calculated
(i) by taking into account the average measurement uncertainties
and time sequences of the individual surveys;
(ii) by assuming circular orbits, 
random orientation, random starting orbital phase,
a mass ratio chosen randomly between 0.2 and 1; and 
(iii) by applying a threshold for variability detection of $p < 10^{-3}$.
Figures\,\ref{fig:vgl_surveys1}--\ref{fig:vgl_surveys2} show
the result. 
The orbital separation range that can be probed with 50\% detection efficiency
is $\leq$3\,AU (subsample of BD/VLMS in this work), 
$\leq$0.6\,AU (Guenther \& Wuchterl 2003), 
$\leq$0.3\,AU (Kurosawa et al. 2006), and 
$\leq$0.2\,AU (Basri \& Reiners 2006),
respectively.
The difference in the maximum orbital separations 
between the surveys by Kurosawa et al. (2006), Guenther \& Wuchterl (2003),
and this work can be largely attributed to differences in the
covered time base. 
While the distribution of the detection efficiency in the survey by Basri \& Reiners (2006) is non-zero 
for separations greater than 1\, AU due to the relatively large time base,
it is decreasing more rapidly compared to the other surveys.
This is mainly due to the moderate RV precision of this survey, which makes it less sensitive to
low-mass ratio systems. 
The differences of this calculated detection probability curve for the survey by Basri \& Reiners (2006)
compared to the one presented by the authors themselves, who find a sensitivity up to 6\,AU,
are the assumptions concerning
inclination (random orientation instead of fixed inclination), mass ratio ($q$ selected randomly from
a uniform distribution between 0.2 and 1 instead of $q$=0.5), and threshold for variability 
($p < 10^{-3}$, i.e. 3.3$\sigma$ instead of $p < 10^{-1.347}$, i.e. 2$\sigma$). 
This, in addition to neglecting differences in the RV errors of different
surveys, leads to a deviating picture in Fig.\,4 of Basri \& Reiners (2006).

\begin{figure*}
\centering
\includegraphics[width=\linewidth,angle=0]{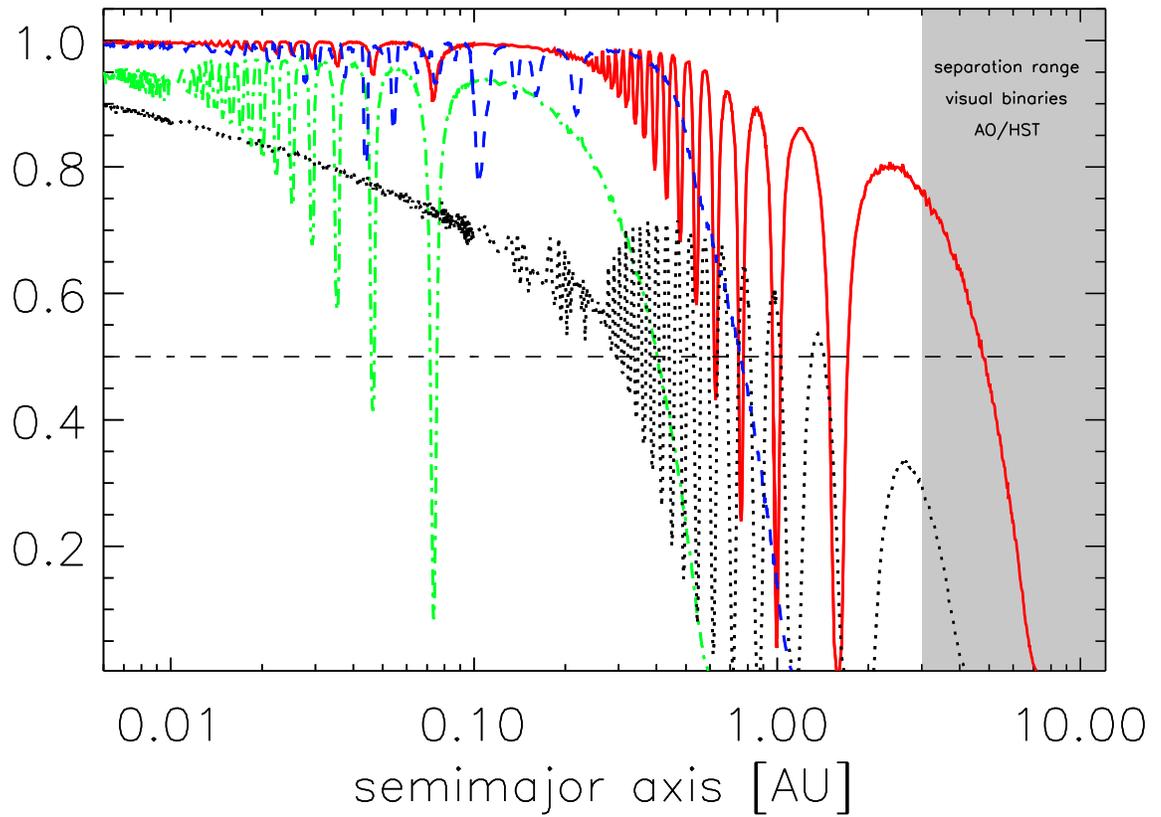}
\caption{
\label{fig:vgl_surveys3}
{\bf Comparison of detection probability of different BD/VLMS RV surveys for high-mass ratio binaries.} 
Same as Fig.\,\ref{fig:vgl_surveys1} 
but for $q$ selected randomly from [0.8,1].
See online version for a color figure.
}
\end{figure*}

\begin{figure*}
\centering
\includegraphics[width=\linewidth,angle=0]{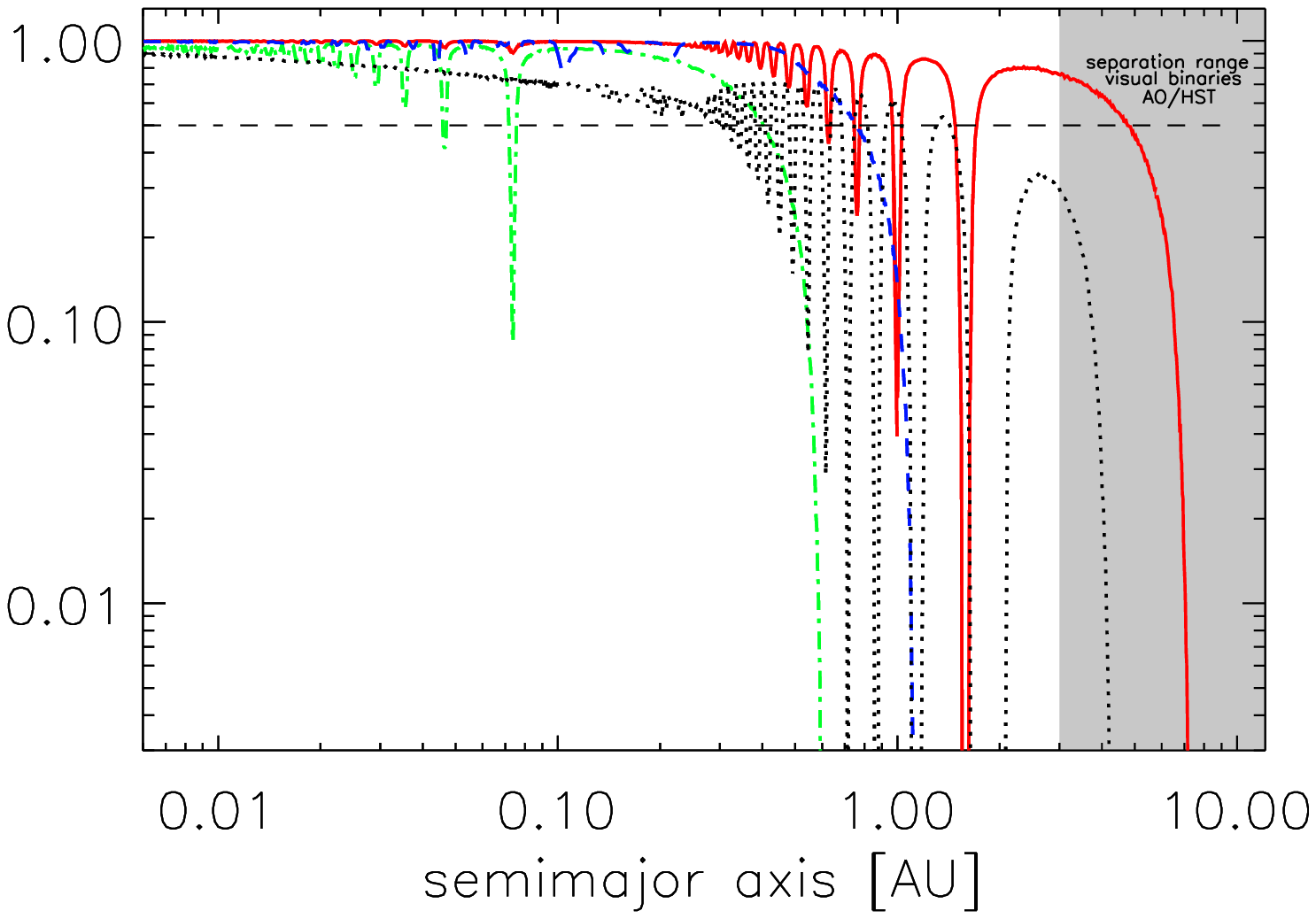}
\caption{
\label{fig:vgl_surveys4}
{\bf 
Same as Fig.\,\ref{fig:vgl_surveys3} but with logarithmic y-axis
to facilitate comparison with other publications.}
See online version for a color figure.
}
\end{figure*}

The BD/VLM binaries with a mass ratio close to unity cause a stronger RV signal
compared to low-mass ratio systems
and are, therefore, easier detected at larger separations.
To quantify this, the sensitivity of the considered surveys 
has been calculated for binaries with mass ratios
between 0.8 and 1. 
In this case ($q>$0.8),
a 50\% detection efficiency can be read off from 
Figs.\,\ref{fig:vgl_surveys3} and \ref{fig:vgl_surveys4}
for the following separation ranges:
$\leq$5\,AU (subsample of BD/VLMS in this work), 
$\leq$2\,AU (Basri \& Reiners 2006),
$\leq$0.8\,AU (Guenther \& Wuchterl 2003), and 
$\leq$0.4\,AU (Kurosawa et al. 2006).
In this mass-ratio regime, neglecting line-blending 
in the performed simulation is relevant (see above).
At separations where the effects of line blending in the $q$ range 0.8-1.0
are significant, these detection probabilities may be significantly reduced.
A comprehensive analysis of these effects is beyond the scope of this
work.

For eleven objects in the survey by Basri \& Reiners (2006), 
there are a third or more epoch observations
with a temporal separation of more than one day.
Such intermediate observations can positively influence the
detection efficiency, as shown before. However, 
a simulation assuming an average observing pattern of $t_{obs}$=[0,500\,d,1650\,d]
for this survey 
does not change the calculated detection
probability.

\section{Low-mass, long-period spectroscopic binary CHXR\,74}
\label{sect:chxr74}

Significant RV variability indicating the presence of a spectroscopic companion
orbiting the very young low-mass star CHXR74 (M4.25-M4.5,$\sim$0.2--0.25\,M$_{\odot}$,
Comer\'on et al. 2000; Luhman 2007) has been already found by
Joergens (2006a). The new RV 
measurements of CHXR\,74 in 2005 and 2006 (Table\,\ref{tab:rvs}) support this finding.
The RV of this star is, apart from some short-term scatter, 
increasing almost monotonically between 2000 and 2006
(Fig.\,\ref{fig:rvs4}, bottom panel) with a recorded 
peak-to-peak difference of 4.8\,km\,s$^{-1}$.
The data indicate a relatively long orbital period of 
at least twelve years, hence,
an orbital separation greater than about 3\,AU.
This can be used to derive a very rough estimate  
of the minimum companion mass $M_2 \sin i$ of $\sim$80\,$\mjup$. 
Given the relatively long orbital period for a spectroscopic system
and its youth,
CHXR\,74 might be one of the first very young 
spectroscopic binaries that can be spatially resolved.
A binary with a separation of more than 3\,AU (20 milli arc sec at the distance of the target) 
might be spatially distinguished from the primary by high-resolution 
imaging, e.g. with NACO/VLT.
If so, CHXR\,74 might provide 
dynamical mass determinations of its components and, therefore, valuable constraints
for the early evolution at the low-mass end of the stellar mass spectrum. 

%
%
%
%
%

\section{Summary and discussion}
\label{sect:disc}

This paper presents an RV survey for close companions
to very low-mass objects in the Cha\,I star-forming region
based on UVES/VLT spectra.
The survey is sensitive to companions 
at orbital distances of 3\,AU and smaller (averaged over all objects) with a
detection rate of 50\% or more, as shown by a
Monte-Carlo simulation. 
For BD/VLM binaries with a mass ratio close to unity ($q>$0.8),
the survey is sensitive to even larger separations.
This is a significant extension to larger orbital separations
compared to previous results of this survey
(Joergens 2006a), as well as compared to other RV surveys of young BD/VLMS 
(Kurosawa et al 2006; Prato 2007a; Maxted et al. 2008), which cover 
separations $\lesssim$0.3\,AU, and to other surveys of field BD/VLMS
(Reid et al 2002; Guenther \& Wuchterl 2003; Basri \& Reiners 2006), 
which cover separations $\lesssim$0.6\,AU (comparison based on $q$ between 0.2 and 1, 3.3\,$\sigma$ detection,
50\% detection probability).
Only the survey by Basri \& Reiners (2006) has a comparable time base, however,
the survey presented here is, nevertheless, superior in terms of detection efficiency  
mainly because of smaller RV errors.
Thus, for the first time, the binary frequency of BD/VLMS 
is probed for the whole separation range $<$3\,AU with a sensitivity to low-mass ratio systems.

Two objects 
were identified as spectroscopic binaries based on significant RV variability.
This corresponds to an observed binary fraction of
18$^{+20}_{-12}$\,\% for the whole sample of eleven BDs and 
(very) low-mass stars (M$\lesssim 0.25\,\msun$)
and 10$^{+18}_{-8}$\,\% for the subsample of ten BD/VLMS 
(M$\lesssim 0.1\,\msun$), respectively.
The detected binaries are
(i) the BD/VLMS \chaha8 (M5.75-M6.5), which
has a substellar RV companion
in a $\sim$1\,AU orbit, as previously discovered
(Joergens 2006; Joergens \& M\"uller 2007). \chaha8
is likely the lowest mass RV companion found so far
around a BD/VLMS.
(ii) The low-mass star CHXR\,74 (M4.25-M4.5), which has a spectroscopic companion
in a relatively long-period orbit (presumably $>$12~years), as found by previous observations
(Joergens 2006) and confirmed here based on new RV data.
With such a relatively long period,
CHXR\,74 is coming into reach for being one of the first very young 
low-mass spectroscopic binaries that can be spatially resolved. 

These spectroscopic binaries appear to have 
orbital periods of at least a few years, i.e.
orbital separations of 1\,AU or greater: 
(i) an RV orbit solution shows that
\chaha8 has a companion in an orbit of more than four years (Joergens \& M\"uller 2007);
(ii) the companion of CHXR\,74 has very likely an orbit of more than twelve
years (Sect.\,\ref{sect:chxr74}).
Thus, while the rate of BD/VLM binaries at $\leq$3\,AU is found
to be 10\% (18\% including CHXR\,74),
at separations $<$1\,AU, it might be less than 10\%
(0/11 for the whole sample and 0/10 for the BD/VLMS subsample). 
There were no signs of the presence of 
shorter period companions around the targets.
This is noteworthy because those cause 
stronger RV signals making them 
easier to detect. In fact, two out of the four BD/VLM spectroscopic
binaries with determined orbital parameters (including \chaha8) 
have periods of only a few days
(Basri \& Mart\'{\i}n 1999; Stassun et al. 2006). 

There is no evidence in the cross correlation peaks of
a double-lined nature of the detected binaries.
This hints at a low rate of double-lined spectroscopic binaries in the sample.

In this survey, there are no large velocity amplitudes recorded (Fig.\,\ref{fig:deltaRV}), 
as would have been expected for (nearly) equal mass binaries
(e.g. a few tenths of km\,s$^{-1}$ for the BD binary PPl\,15, Basri \& Mart\'{\i}n 1999). 
The detected significant RV variability occurs only on small scales: 
\chaha8 has an RV semi-amplitude of 1.6\,km\,s$^{-1}$ (Joergens \& M\"uller 2007) and a
half peak-to-peak RV difference for CHXR\,74 is 2.4\,km\,s$^{-1}$.
Note that an only moderate RV precision of e.g. 1.4\,km\,s$^{-1}$, 
as in the survey by Basri \& Reiners (2006), 
would have prevented the detection of these binaries.
The results hint at the possibility that there exist more low-mass ratio systems 
among BD/VLMS, at least at $<$3\,AU, than anticipated from an extrapolation from
the mass-ratio distribution derived based on 
direct imaging surveys, which peaks at $q\sim$1 and declines rapidly for low $q$ values
(e.g. Burgasser et al. 2007).
On the other hand, the small RV amplitudes can also be attributed to 
observations at low orbital inclinations.
It is, however, interesting that 
a preference for small RV amplitudes is also seen in RV surveys
of BD/VLMS in other star-forming regions, namely in USco and $\rho$\,Oph 
(0.2--0.5\,km\,s$^{-1}$, Kurosawa et al. 2006, cf. also footnote 5)
and in Taurus, Ophiuchus, and TW Hya (1.5--2.0 km\,s$^{-1}$, Prato 2007a).

Current spectroscopic surveys of very low-mass objects differ widely 
in the studied parameter ranges, as
orbital separations, (primary and) companion masses, and environments/ages.
Within the (sometimes large) statistical errors, the determined binary frequencies are 
largely consistent and range between $<$7.5\% and 24\%: 
10$^{+18}_{-8}$\,\% for ten BD/VLMS in Cha\,I at $\leq$3\,AU (this work),
24$^{+16}_{-13}$\,\% for 17 BD/VLMS in USco and $\rho$\,Oph at $\leq$0.3\,AU
(Kurosawa et al. 2006)\footnote{The four objects classified 
as spectroscopic binaries in the survey by Kurosawa et al. (2006) display 
very small RV differences (0.2--0.5\,km\,s$^{-1}$).
It has to be further investigated whether it can be excluded that they are caused by systematic errors.
If interpreted as RV semi-amplitudes caused by companions at $\leq$0.1\,AU, these companions
would have masses M$_2 \sin i$ in the planetary regime (1--2\,$\mjup$)
(see Sect.\ref{sect:RVvari}). In both cases, systematic errors or planets, 
we would not count them as binaries yielding a 
binary fraction of this survey of $\leq6\%$ (0/17).},
17\% for 18 BD/VLMS in Taurus, Ophiuchus, and TW Hya at $\lesssim$0.3\,AU
(Prato 2007a), 
$<$7.5\% for 51 BD/VLM member candidates of $\sigma$ and $\lambda$
Orionis at $\leq$0.3\,AU
(Maxted et al. 2008), 
12$^{+11}_{-7}$\,\% for 25 field and cluster dwarfs at $\leq$0.6\,AU (Guenther \& Wuchterl 2003),
and 11$^{+7}_{-4}$\,\% for 53 (very) low-mass field stars and BDs
at $\leq$0.2\,AU ($\leq$1-2\,AU for $q>$0.8) (Basri \& Reiners 2006).

By combining my work in Cha\,I with other RV surveys of young regions,
an overall BD/VLM binary fraction at an age of a few million years can be derived. 
With the exception of Cha\,I, all RV surveys of young BD/VLMS
probe only a narrow separation range ($\lesssim$0.3\,AU). 
No companions are found orbiting BD/VLMS in Cha\,I at these 
separations.
Thus, the overall observed binary fraction at separations $\leq$0.3\,AU
of very young BD/VLMS 
in Cha\,I (0/10), USco and $\rho$\,Oph (4/17, Kurosawa et al. 2006), 
Taurus, Ophiuchus, and TW~Hya (3/18, Prato 2007a), and 
$\sigma$ and $\lambda$~Orionis (0/52, Maxted et al. 2008)
 is 7$^{+5}_{-3}$\% (7/97).

Comparing the binary frequency of BD/VLMS in ChaI
for separations $\leq$3\,AU (10$^{+18}_{-8}$\,\%) with what is
determined based on direct (adaptive optics and HST) imaging surveys at 
larger separations ($>$20\,AU) \emph{in the same star-forming region}
(9$^{+17}_{-7}$\%, Neuh\"auser et al 2002; Neuh\"auser, Guenther \& Brandner 2003; 
11$^{+9}_{-6}$\%, Ahmic et al 2007),
one finds consistent values. 

\section{Conclusions}
\label{sect:concl}

For the first time, the binary frequency of BD/VLMS is probed
in the orbital separation range 1--3\,AU.
The separation distribution of BD/VLM binaries detected by direct imaging surveys
(e.g. Burgasser et al. 2007)
has a peak around 3--10\,AU, which is close to the incompleteness limit of these surveys
($<$2--3\,AU for the field and $<$10-15\,AU for star-forming regions, respectively).
It seemed, therefore, possible that the largest fraction of BD/VLM binaries has
separations in the range of about 1--3\,AU and still remained undetected.
The BD/VLM binary frequency of 10$^{+18}_{-8}$\,\%
determined here at separations
of 3\,AU and smaller shows that this is not the case because the rate of 
spectroscopic binaries at $<$3\,AU does not significantly exceed
the rate of resolved binaries at larger separations 
(10--30\%, e.g. Burgasser et al. 2007).
Thus, the separation range $<$3\,AU missed by direct
imaging surveys is not adding a significant fraction to the BD/VLM binary frequency,
and the overall binary frequency of BD/VLMS is in the range 10-30\%.
The finding of no signs (0/10, i.e. $\leq$10\%) for companions at $<$1\,AU
is consistent with a decline in the separation distribution of BD/VLM binaries
towards smaller separations starting between 1 and 3\,AU.
Furthermore, the previous finding that the observed mass-dependent decrease in the stellar binary frequency
(e.g. Delfosse et al. 2004; Shatsky \& Tokovinin 2002; Kouwenhoven et al. 2005)
continuously extends into the BD regime (e.g. Burgasser et al. 2007)
is confirmed here for separations $<$3\,AU.
This is consistent with a continuous formation mechanism from 
low-mass stars to BDs.
 
These results are based on a relatively small sample.
By combining this work in Cha\,I with surveys of other young regions,
which are restricted to separations $\leq$0.3\,AU,
an overall observed binary fraction of very young 
BD/VLMS can be determined with better statistics but with a more limited
separation range. In this way, a binary frequency of 7$^{+5}_{-3}$\% (7/97) is found
at $\leq$0.3\,AU.
In the near future, the sample of the survey in Cha\,I will be substantially 
enlarged, allowing the re-investigation of the 
binary frequency at $<$3\,AU of BD/VLMS in Cha\,I with significantly improved statistics
for a uniformly obtained data set. 
Current and future observational efforts are, in addition, directed towards
determination of orbital parameters of the detected binary systems, 
including RV follow-up and high-resolution imaging/astrometry.

\clearpage
\begin{table*}
\begin{minipage}[t]{\columnwidth}
\begin{center}
\caption{
\label{tab:rvs} 
{\bf New RV measurements for BDs and VLMSs in Cha\,I.}
}
\renewcommand{\footnoterule}{}  
\begin{tabular}{lcccl}
\hline
\hline
\myrule
Object               &   Date      & HJD           & RV             & ~$\sigma_{RV}$\footnote{Error of the \emph{relative} RVs. 
In most cases (marked with an asterisk) this is the sample standard deviation derived from
two consecutive measurements.} \\
                     &             &               & [km\,s$^{-1}$] & [km\,s$^{-1}$] \\
\hline
\myrule

Cha\,H$\alpha$\,1    & 2006 Jan 16 & 2453751.74004 & 16.206 &  0.39\,* \\

\hline
Cha\,H$\alpha$\,2    & 2006 Jan 16 & 2453751.79554 & 16.319 & 0.25\,* \\

\hline
Cha\,H$\alpha$\,3    & 2006 Jan 15 & 2453750.81659 & 16.339 & 0.70\,* \\

\hline
Cha\,H$\alpha$\,5    & 2006 Jan 15 & 2453750.84774 & 15.697 & 0.03\,* \\

\hline
Cha\,H$\alpha$\,6    & 2006 Jan 16 & 2453751.82994 & 16.630 & 0.07\,* \\

\hline
Cha\,H$\alpha$\,7    & 2006 Jan 15 & wrong pointing \\

\hline
Cha\,H$\alpha$\,12   
		     & 2006 Jan 17 & 2453752.70145 & 15.831 & 0.05\,* \\

\hline
B\,34                & 2005 Nov 25 & 2453699.83976 & 15.687 & 0.10\,*  \\

\hline

CHXR\,74             & 2005 Mar 21 & 2453450.58923  & 18.185 & 0.06\,* \\

                     & 2006 Apr 10 & 2453835.63456  & 17.982 & 0.10 \\

                     & 2006 Jun 15 & 2453901.52267 & 19.063 & 0.10 \\

\hline
\end{tabular}
\end{center}
\end{minipage}
\end{table*}

\begin{table*}
\begin{minipage}[t]{\columnwidth}
\begin{center}
\caption{
\label{tab:chi2prob} 
{\bf Characteristics of observations and $\chi^{2}$ probabilities.}}
\renewcommand{\footnoterule}{}  
\begin{tabular}{lccccccc}
\hline
\hline
\myrule
Object & $\Delta$ RV\footnote{half peak-to-peak RV differences} & $\overline\mathrm{RV}_{weighted}$\footnote{
the error of the weighted mean takes an uncertainty of 400\,m\,s$^{-1}$ into account for the
zero point of the velocity} & no. obs\footnote{includes data from Table\,\ref{tab:rvs}, Joergens (2006a), and 
Joergens \& M\"uller (2007)} & Min. $\Delta$t & Max. $\Delta$t & \em{p} & \\
                     & [km\,s$^{-1}$] & [km\,s$^{-1}$] & & [days] & [days] &\\
\hline
\myrule

Cha\,H$\alpha$\,1    & 0.24 & 16.27 $\pm$ 0.52 & 3 & 20 & 2113 & 8.3 $\times 10^{-1}$ \\

\hline
Cha\,H$\alpha$\,2    & 0.15 & 16.25 $\pm$ 0.49 & 3 & 20 & 2113 & 8.7 $\times 10^{-1}$ \\

\hline
Cha\,H$\alpha$\,3    & 0.99 & 14.86 $\pm$ 0.89 & 3 & 19 & 2111 & 5.6 $\times 10^{-2}$ \\

\hline
Cha\,H$\alpha$\,4    & 0.21 & 14.85 $\pm$ 0.43 & 11 & 1 &  701 & 1.0  \\

\hline
Cha\,H$\alpha$\,5    & 0.13 & 15.59 $\pm$ 0.48 & 3 & 19 & 2111 & 8.7 $\times 10^{-1}$ \\

\hline
Cha\,H$\alpha$\,6    & 0.28 & 16.52 $\pm$ 0.56 & 3 & 19 & 2112 & 6.3 $\times 10^{-1}$ \\

\hline
Cha\,H$\alpha$\,7    & 0.58 & 17.09 $\pm$ 0.98 & 2 & 19 &   19 & 1.5 $\times 10^{-1}$ \\

\hline
Cha\,H$\alpha$\,8    & 1.38 & 16.62 $\pm$ 0.67 & 11 & 3 & 2542 & 0.0 & vari. \\

\hline
Cha\,H$\alpha$\,12   & 0.96 & 15.29 $\pm$ 0.93 & 3 & 20 & 2113 & 5.6 $\times 10^{-3}$ \\
		     
\hline
B\,34                & 0.07 & 15.74 $\pm$ 0.42 & 6 & 6 & 2083 & 1.0  \\

\hline
CHXR\,74             & 2.39 & 16.71 $\pm$ 0.81 & 13 & 1 & 2285 & 0.0 & vari. \\

\hline
\end{tabular}
\end{center}
\end{minipage}
\end{table*}

\clearpage


\clearpage
\clearpage
\clearpage

\clearpage
\clearpage

\clearpage

\begin{acknowledgements}
I would like to thank C. Bailer-Jones, W. Brandner, Th. Henning,
R. Kurosawa, S. More 
for fruitful discussions
and an anonymous referee for helpful comments that improved the 
paper.
Further, I like to acknowledge the excellent work of the 
ESO staff at Paranal, who took the UVES spectra of this paper
in service mode. 
\end{acknowledgements}

\end{document}